\documentclass[%
 reprint, 
 amsmath,amssymb,
 aps, 
prb,
]{revtex4-2}
\usepackage{svg}
\usepackage{amssymb}
\usepackage{booktabs}
\usepackage{graphicx}
\usepackage{dcolumn}
\usepackage{bm}

\begin{document}

\preprint{APS/123-QED}

\title{\textbf{Fast Single Nitrogen-Vacancy Center Ramsey Characterization using a Physics-Informed Neural Network} 
}%

\author{Chao Shang}
 \email{cs2485@cornell.edu}
\author{Gregory D. Fuchs}%
 \email{gdf9@cornell.edu}
\affiliation{%
 School of Applied and Engineering Physics, Cornell University, Ithaca, New York 14853, USA 
}%

\date{\today}
\begin{abstract}
Precise characterization of the local spin environment of single diamond nitrogen-vacancy (NV) centers is crucial for advancing quantum sensing, quantum networking, and the optimization of quantum materials. However, single NV center fluorescence measurements requires long averaging times to obtain clean data that is suitable for conventional model fitting, and that constitutes a key experimental bottleneck for high-throughput characterization. To address this, we introduce \textsc{NVRNet}, a physics-informed simulation-to-reality machine learning pipeline that maps minimal-sweep, noisy Ramsey data to a denoised waveform while directly estimating the hyperfine coupling to proximal ${}^{13}\mathrm{C}$ nuclear spins. The pipeline's denoiser utilizes a two-stage time--frequency U-Net and an attention-augmented time-domain U-Net, pretrained on Hamiltonian-based spin-dynamics simulations with experimentally calibrated noise. To effectively bridge the simulation-to-reality gap, parameter-efficient adapters are attached to the backbone and fine-tuned on targeted experimental data. Across three distinct NV centers, this experimentally fine-tuned model reduces the median reconstruction error on held-out, few-sweep traces to $0.44\text{--}0.67\times$ of the raw experimental noise level. Subsequently, a transformer-based estimator extracts the underlying hyperfine parameters. Forward reconstructions derived from these inferred parameters faithfully reproduce the dominant experimental time- and frequency-domain features, yielding representative normalized fast Fourier transform (FFT) reconstruction errors of $0.10\text{--}0.19$. By reducing both the required data volume and acquisition time, \textsc{NVRNet} enables up to $\sim 40\times$ acceleration of the measurement process, establishing a fast, hardware-compatible pathway for robust hyperfine inference and autonomous qubit characterization.
\end{abstract}

\maketitle



\section{Introduction}

Nitrogen-vacancy (NV) centers in diamond are a solid-state spin qubit system with room-temperature optical initialization and readout, microwave manipulation, and long spin coherence, making them a versatile platform for quantum sensing and quantum information science \cite{Doherty2013PhysRep,Jelezko2006PSSA,Awschalom2018NatPhoton,Degen2017RMP}. Applications span nanoscale vector magnetometry (single-NV scanning probes and wide-field ensembles) \cite{Maze2008Nature,Taylor2008NatPhys,Maletinsky2012NatNano,Grinolds2013NatPhys,Rondin2014RPP,Tetienne2018PRB}, 
biosensing with fluorescent nanodiamonds \cite{Schirhagl2014ARPC,McGuinness2011NatNano}, rotation sensing via geometric-phase protocols \cite{soshenko2021nuclear,76pr-366b,jarmola2021demonstration,PhysRevLett.133.150801,ajoy2012stable}, and nuclear-spin--assisted quantum registers, with strain and electromechanical coupling providing routes to hybrid transduction \cite{Childress2006Science,Taminiau2012PRL,Taminiau2014NatNano,MacQuarrie2015Optica,Ovartchaiyapong2014NatCommun,Teissier2014PRL}. 

A practical bottleneck is the rapid and reliable characterization of individual NV centers and their local nuclear spin environments \cite{Hopper2018Micromachines,Awschalom2018NatPhoton}. 
Standard workflows include pulsed measurement protocols (e.g. Rabi, Ramsey, spin echo) to extract photoluminescence (PL) contrast, $T_2^\ast$, and $T_2$, and information about the local spin environment such as the hyperfine coupling to proximal $^{13}\mathrm{C}$ spins \cite{Doherty2013PhysRep,Rondin2014RPP,Taminiau2012PRL, Taminiau2014NatNano}. 
In the short data acquisition time regime, where the signal amplitude is near the shot-noise limit, weak hyperfine beating is difficult to resolve and least-squares model fitting often fails to disambiguate the start of beating from real dephasing. Therefore, one typically requires long time averaging or longer dynamical-decoupling protocols which reduces the throughput for screening \cite{Degen2017RMP,Rondin2014RPP,Hopper2018Micromachines}. Spectral crowding from small $^{13}\mathrm{C}$ couplings can lead to overlapping signatures that are easily masked by the experimental artifacts like drift and sampling misalignment \cite{Taminiau2012PRL,Taminiau2014NatNano,Doherty2013PhysRep}. Fitting-based analysis is tricky when there are multiple environmental spins interacting with the NV center, thus requiring significant decision-making by the experimentalist to establish the details of the model~\cite{Hopper2018Micromachines}.%

These challenges motivate an exploitation of machine-learning methods tailored to NV center-related time-series data. In broader signal-processing domains—such as speech enhancement, audio denoising, and general nonstationary time-series regression: 1D UNet encoder–decoder architectures are effective because they combine multiscale feature extraction with skip connections that preserve localized structure while suppressing baseline drift and broadband noise~\cite{ronneberger2015uUNet,ho2020denoisingUnet,wu2023metaattentionUnet,drozdzal2016importanceUnetSkip,azad2022contextualattentionUnet}. Attention-based models provide a complementary mechanism: by allowing global interactions across a sequence, self-attention can disentangle multiple, simultaneously present spectral components and track long-range correlations that are difficult to represent with purely local convolutions~\cite{vaswani2017attention,DBLP:journals/corr/BahdanauCB14,shih2019temporalattention,azad2022contextualattentionUnet}.

Recent work on deep learning in NV spectroscopy has primarily adopted CNN-based encoder–decoder designs and focused on multi-pulse “fingerprint” sequences\cite{varona2024automaticCPMG,jung2021deepcpmg,xu2023noise}. While these architectures perform well for signals dominated by repetitive motifs, such as a sequence of dips with similar morphology. However,they are less well suited to the more intricate, multi-frequency oscillatory structure typical of Ramsey traces in complex hyperfine environments. Moreover, much of the reported parameter-estimation capability has been demonstrated predominantly on simulated datasets, leaving two practical gaps for deployment: 1. \emph{Incomplete coverage of real laboratory noise mechanisms}, such as laser intensity drift or slow mechanical drift of the objective/sample relative position. 2. \emph{Limited explicit treatment of measurement uncertainty.} As a result, models trained on idealized simulations can generalize poorly to experimental data, motivating approaches that reduce pulse complexity and acquisition time while explicitly addressing long-range structure and simulation-to-reality mismatch~\cite{varona2024automaticCPMG,jung2021deepcpmg}.

\begin{figure}[b]
\includegraphics[width=0.5\textwidth,]{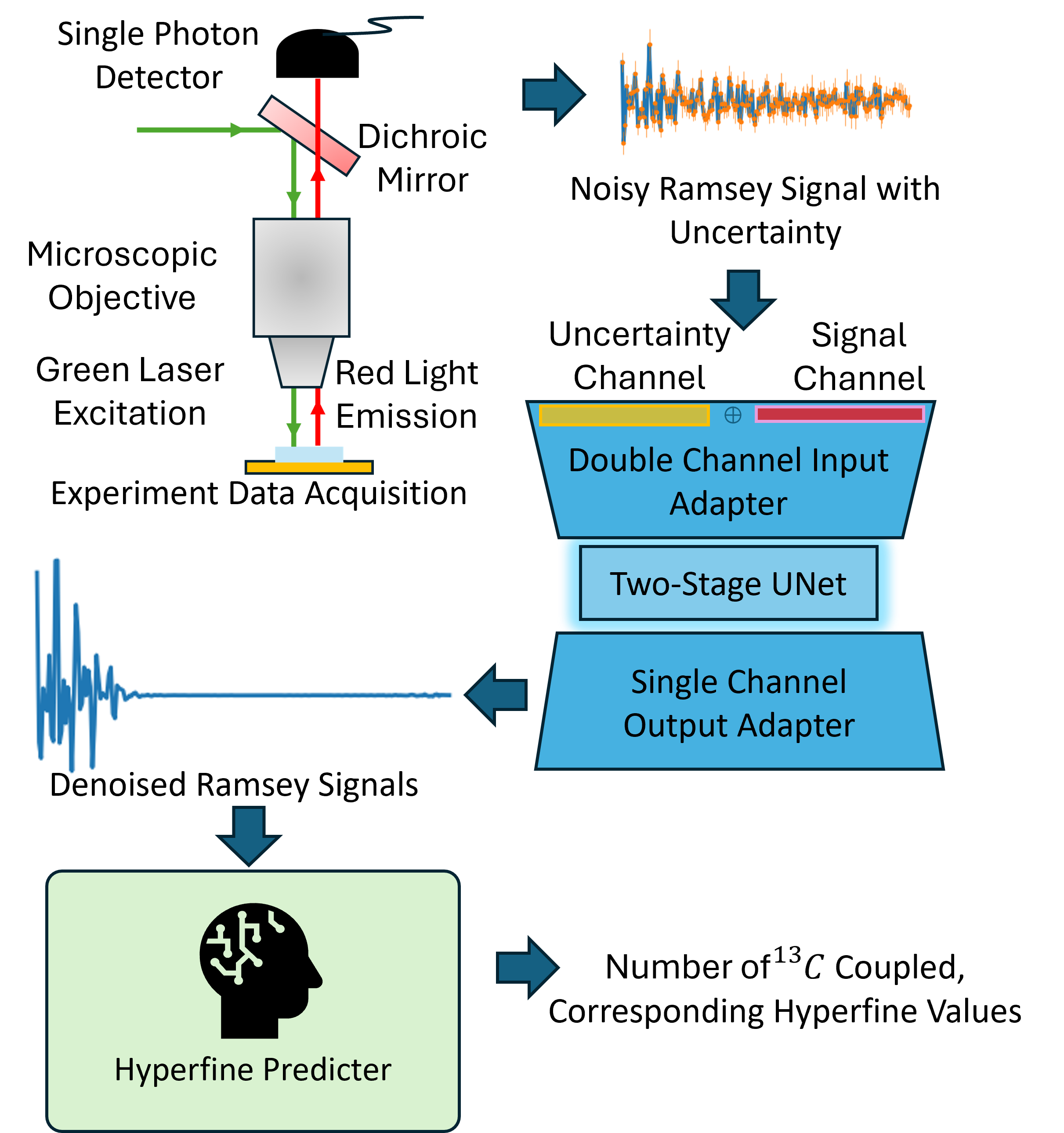}
\caption{\label{workflow}
\textbf{Workflow of \textsc{NVRNet}.}
A Ramsey PL trace is collected under room temprature with single photon detector. Then it is first processed by the denoising module, which comprises a simulation-pretrained core augmented with uncertainty-aware adapter layers, fine-tuned on experimental traces. The resulting denoised trace is then passed to the parameter-estimation network, which predicts the ${}^{13}\mathrm{C}$ count (up to $n_{\max}=9$) and the associated parallel hyperfine couplings $A_{\parallel}$.}

\end{figure}

This work addresses these challenges by targeting the minimal averaging regime: a few sweeps of noisy Ramsey traces (with $\sim10^{3}$ photons collected for each free-induction-decay period measurement) that still contain hyperfine information in weak beating components. We introduce \textsc{NVRNet}, a physics-informed machine-learning pipeline that converts noisy Ramsey PL contrast time series data into (i) a denoised trace suitable for conventional fitting and (ii) a direct estimate of the hyperfine parameters, including the ${}^{13}\mathrm{C}$ count (up to $n_{\max}=9$) and parallel couplings. The workflow is shown in Fig.~\ref{workflow}. The central advance is a principled simulation-to-reality strategy that combines large-scale simulation pretraining with lightweight experimental adaptation. A two-stage time--frequency denoiser is pretrained on synthetic Ramsey signals generated from a realistic NV Hamiltonian model with randomized nuclear configurations and experimentally calibrated noise statistics. Although this simulation pipeline is explicitly tuned to reproduce the experimentally observed residual structure, it cannot perfectly capture all laboratory-specific, nonstationary noise processes and, crucially, does not intrinsically encode how the mapping should adapt as the input uncertainty level changes. To address this, we adopt a pretraining--post-training strategy in the same spirit as modern large-language-model pipelines: we freeze the pretrained backbone and fine-tune only uncertainty-aware adapter layers on a small set of experimental traces, providing a low-parameter mechanism to learn experimental artifacts without catastrophic overfitting \cite{ouyang2022traininglanguagemodelsfollow,Guo_2025}.
 The denoised output is then fed to a dedicated parameter-estimation network trained on simulation, which predicts the hyperfine content and is validated experimentally by \emph{forward reconstruction}: inferred parameters are mapped back through the physical model to reproduce the observed time- and frequency-domain features.

By explicitly separating (i) a robust, transferable denoising stage optimized for the few-sweep regime and (ii) a physics-constrained parameter estimator trained with deterministic labels, \textsc{NVRNet} tackles two long-standing obstacles in automated NV characterization: the sensitivity of inference to realistic experimental noise and the human time spent initializing and stabilizing hyperfine fits. We demonstrate that adapter fine-tuning consistently improves denoising accuracy and robustness on held-out experimental traces, and that the parameter estimator recovers hyperfine-relevant spectral structure sufficient to provide reliable initialization for downstream refinement. This enables a substantially faster path from raw Ramsey data to actionable hyperfine information, supporting high-throughput screening and more scalable calibration of NV-based quantum devices.

\section{\label{sec:level1}Physical Modeling and Data Generation}
We construct a physics-grounded synthetic dataset by first generating stochastic realizations of an NV center’s local nuclear-spin environment from explicit diamond lattice geometry, extracting hyperfine parameters directly from the resulting spatial configuration. Using these parameters, we simulate Ramsey free-induction signals from a reduced rotating-frame NV Hamiltonian and subsequently apply an experimentally calibrated noise model to produce realistic noisy training data.

\subsection{\label{sec:lattice}Lattice Structure Generation}
For each realization, we generate a finite diamond supercell, form an NV center via a vacancy–nitrogen substitution, and stochastically dope ${}^{13}\mathrm{C}$ at natural abundance, retaining only spins within a cutoff radius $R_c=6~\mathrm{\AA}$ to form a compact bath. Hyperfine couplings are assigned using the secular dipolar scaling $C_i \propto r_i^{-3}(3z_i^2/r_i^2-1)$; full construction details, cutoff justification, and ${}^{13}\mathrm{C}$ statistics are provided in Appendix~\ref{LatticeAppendix}.

\subsection{\label{sec:level2}Ramsey Signal Simulation}
For each lattice configuration, we simulate Ramsey free induction in the $\{|m_s=0\rangle,|m_s=-1\rangle\}$ manifold using a rotating-frame model in which perpendicular hyperfine terms are neglected to leading order under $\omega_0\gg A_{\perp}$. We simulate the Ramsey protocol using a semi-quantum dynamical method implemented in the \textsc{Qutip} Package \cite{qutip5}. We average over quasi-static nuclear spin projections, apply an empirical Gaussian $T_2^\ast$ envelope, and map the resulting population to photoluminescence contrast via a linear readout model. Simulation details, approximated Hamiltonian derivation, and averaging details are provided in Appendix~\ref{app:rwa_nv}.

\subsection{\label{sec:noisemodeling}Noise Modeling}
To calibrate the noise model used for large-scale synthetic pretraining, we analyze experimental Ramsey traces. Throughout this work, the measured photoluminescence observable is reported as a normalized readout rather than as raw photon counts:
\begin{equation}
\mathrm{PL}(t)\equiv 100\times \frac{S(t)}{N(t)},
\end{equation}
where $S(t)$ and $N(t)$ denote the signal-window and normalization-window photon counts, respectively. Thus, values reported in PL(\%) correspond to percentage-scaled normalized PL readout, while differences between traces are reported in percentage points of normalized PL.

For each NV center, we construct a high-SNR reference trace $\mathrm{PL}_{\mathrm{ref}}(t)$ from repeated measurements under nominally identical conditions and use it as a practical proxy for the underlying mean Ramsey response. Experimental acquisition, reference-trace construction, and uncertainty-propagation details are provided in Appendix~\ref{DatAug}. For an individual experimental sweep $s$, with sampled evolution times $\{t_k\}$, we define the residual sequence pointwise as
\begin{equation}
r^{(s)}(t_k)\equiv \mathrm{PL}_{\mathrm{exp}}^{(s)}(t_k)-\mathrm{PL}_{\mathrm{ref}}(t_k).
\label{eq:residual_def}
\end{equation}

Across the dataset, the residuals are well described by two dominant contributions: (i) fast fluctuations consistent with shot-noise-limited detection and (ii) slow sweep-to-sweep drift that appears primarily as a baseline offset. Accordingly, we model an experimental trace as
\begin{equation}
\mathbf{y}_{\mathrm{exp}}^{(s)}=\mathbf{y}+\delta^{(s)}\mathbf{1}+\boldsymbol{\epsilon}^{(s)},
\label{eq:noise_vector_model}
\end{equation}
where $\mathbf{y}\in\mathbb{R}^{200}$ denotes the clean reference sequence with entries $y_k\equiv \mathrm{PL}_{\mathrm{ref}}(t_k)$, $\mathbf{1}$ is the all-ones vector, $\delta^{(s)}$ is a trace-level baseline offset, and $\boldsymbol{\epsilon}^{(s)}$ is the fast pointwise noise. The corresponding residual sequence is therefore
\begin{equation}
\mathbf{r}^{(s)}\equiv \mathbf{y}_{\mathrm{exp}}^{(s)}-\mathbf{y}
=\delta^{(s)}\mathbf{1}+\boldsymbol{\epsilon}^{(s)}.
\end{equation}
\emph{Fast heteroscedastic component.}
Conditioned on the clean sequence, we assume independent Gaussian fluctuations at each time point with a variance set by the local PL level,
\begin{equation}
\epsilon^{(s)}_k \text{ is drawn from } \mathcal{N}\!\left(0,\sigma^2(y_k)\right),
\qquad k=1,\dots,200,
\label{eq:hetero_eps}
\end{equation}
and estimate the noise scale function $\sigma(\cdot)$ from data by binning residual entries $r^{(s)}_k$ versus their corresponding reference levels $y_k$ and fitting the empirical standard deviation with a low-order polynomial. In this work we use a quadratic parameterization,
\begin{equation}
\sigma(y)\approx b_0+b_1 y+b_2 y^2,\qquad \sigma(y)>0,
\label{eq:sigmahat}
\end{equation}
with fitted coefficients (in PL units)
\begin{equation}
\sigma(y)\approx -147.9 + 3.481\,y - 0.01975\,y^2,
\label{eq:sigmahat_fit}
\end{equation}
and we clip $\sigma(y)$ to remain positive when sampling synthetic noise.

\emph{Trace-level drift component.}
To capture sweep-to-sweep PL-strength drift, we include a \emph{single} random baseline offset $\delta^{(s)}$ for each synthetic trace (held constant across the 200 time points within that trace). We model this offset as a zero-mean Gaussian variable with variance $\sigma_{\mathrm{dc}}^{2}$,
$\delta^{(s)} $ is drawn from $\mathcal{N}(0,\sigma_{\mathrm{dc}}^2)$, $\sigma_{\mathrm{dc}} = 3.928$ (PL units).

The scale $\sigma_{\mathrm{dc}}$ is estimated empirically from the distribution of per-sweep residual means,
\begin{equation}
\bar{r}^{(s)} \equiv \frac{1}{200}\sum_{k=1}^{200} r^{(s)}_k,
\end{equation}
computed over the experimental sweep ensemble for each NV center.

\paragraph*{Full noise synthesis.}
Given any clean simulated PL sequence $\mathbf{y}=(y_1,\dots,y_{200})$ produced by the physical model, we synthesize a noisy realization by (i) drawing a trace-level baseline offset $\delta$ and (ii) drawing a zero-mean fluctuation sequence $\boldsymbol{\epsilon}=(\epsilon_1,\dots,\epsilon_{200})$ whose pointwise variance is set by the local PL level through Eq.~\eqref{eq:hetero_eps}. We then form the noisy trace as
\begin{equation}
\mathbf{y}_{\mathrm{noisy}}
\;=\;
\mathbf{y}
\;+\;
\delta\,\mathbf{1}
\;+\;
\boldsymbol{\epsilon},
\label{eq:noise_synth}
\end{equation}
where $\mathbf{1}$ is the length-200 all-ones vector. Conditional on $\mathbf{y}$, each entry $\epsilon_k$ is modeled as a zero-mean Gaussian fluctuation with variance $\sigma^2(y_k)$, so the noise amplitude follows the local signal level across the sequence. This vector-valued construction captures both the PL-dependent fast fluctuations and the dominant sweep-to-sweep baseline drift observed in the experimental residual sequences, while remaining simple enough to scale to $\sim 7\times 10^7$ synthetic traces.

\paragraph*{Principal component analysis validation in the time domain.}
To assess whether the simulated dataset reproduces the dominant \emph{structure} of experimentally observed noise and trace-to-trace variability, we perform principal component analysis (PCA) on Ramsey traces that are resampled onto a common time grid and mean-centered \cite{Pearson1901ClosestFit,Hotelling1933PCA}. Denoting the centered data matrix by $\mathbf{X}\in\mathbb{R}^{N\times L}$ (with $N$ traces and $L$ time samples), PCA yields orthonormal temporal modes $\{u_k(t)\}$ and scalar scores $\{c_k\}$ such that an individual trace can be approximated as
\begin{equation}
x(t)\approx \mu(t)+\sum_{k=1}^{K} c_k\,u_k(t),
\label{eq:pca_expand}
\end{equation}
where $\mu(t)$ is the empirical mean trace and $K$ is the number of retained components \cite{Pearson1901ClosestFit}. In our setting each trace contains $L=200$ time samples, so PCA produces up to 200 principal components(PCs); we find that the first two components already capture the majority of the variance, with $\mathrm{PC1}$ and $\mathrm{PC2}$ explaining $\sim 46\%$ and $\sim 13\%$ respectively (cumulatively $\sim 59\%$).

\begin{figure}[b]
\includegraphics[width=0.4\textwidth,]{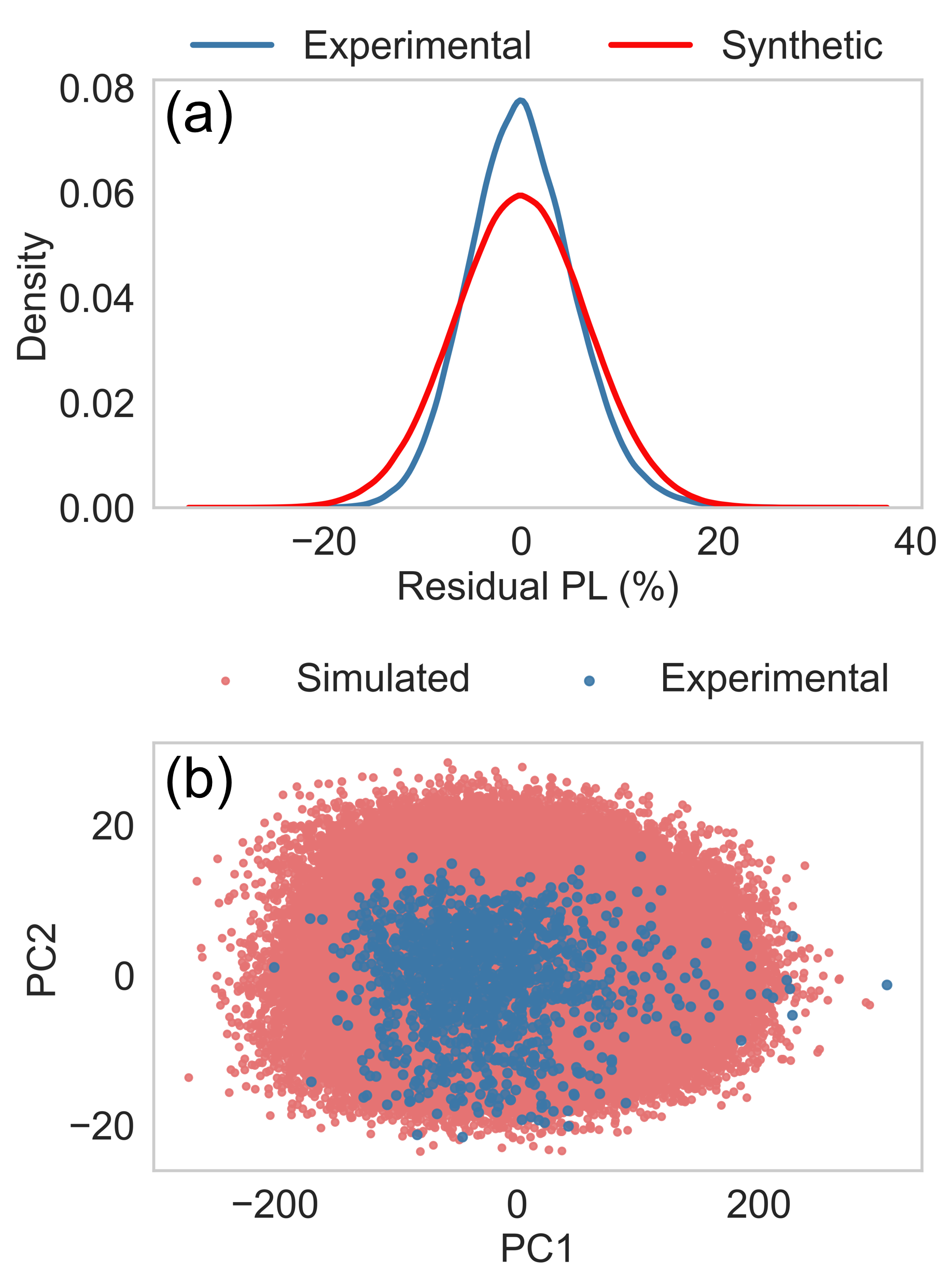}
\caption{\label{fig:pca}\textbf{Comparison between experimental and simulated noise statistics and principal components.}
(a) Distributions in PL(\%) for experimental (blue) and synthetic (pink) residuals.
(b) PCA embedding of full traces: the simulated dataset (pink) spans the region occupied by the experimental data (blue) in the leading two principal components, indicating that simulation covers the dominant experimental variability.
}

\end{figure}

We first compare the marginal distributions of pointwise residuals (experimental and simulated) in Fig.~\ref{fig:pca}(a), showing that the simulation reproduces the overall width and shape of the experimental noise distribution in normalized PL. We next compare the low-dimensional geometry of the two datasets using principal component analysis. Figure~\ref{fig:pca}(b) shows that the simulated traces span the region occupied by the experimental traces in the $(\mathrm{PC1},\mathrm{PC2})$ plane, indicating that the simulation captures the dominant trace-to-trace variability observed in experiment.

Taken together, these results provide a time-domain validation that the synthetic dataset reproduces the principal low-dimensional structure of the experimental measurements, while the remaining mismatch motivates the subsequent adapter-based fine-tuning on experimental traces. The physical interpretation of PC1 and PC2 is discused in Appendix~\ref{app:pca_interp}.

\section{NVRNet}
This section presents \textsc{NVRNet}, a physics-informed learning pipeline that converts few sweep Ramsey PL traces into physically interpretable hyperfine parameters. The workflow is end-to-end: a raw experimental time trace is first denoised to recover the underlying Ramsey oscillation while retaining weak, phase-coherent beating; the denoised output is then encoded jointly in the time and frequency domains and mapped to an estimate of the coupled ${}^{13}$C count and the corresponding parallel hyperfine couplings, which are finally validated by forward reconstruction of the Ramsey response using the same physical model class. We begin by introducing the denoising module, optimized to suppress realistic laboratory noise (shot noise, drift, and contrast fluctuations) without hallucinating oscillatory structure, and we report quantitative and qualitative denoising performance within the same subsection (Sec.~\ref{sec:denoise}). We then describe the supervised hyperfine-parameter estimator trained on labeled simulations, which combines auxiliary ``oscillation-pattern'' regressors with a time--frequency Transformer head, and we evaluate both simulated accuracy and experimental reconstruction fidelity in Sec.~\ref{sec:hf_estimator}.

\subsection{Denoising Network}
\label{sec:denoise}

\subsubsection{\textbf{Model structure}}
\begin{figure*}[t]
\includegraphics[width=7 in]{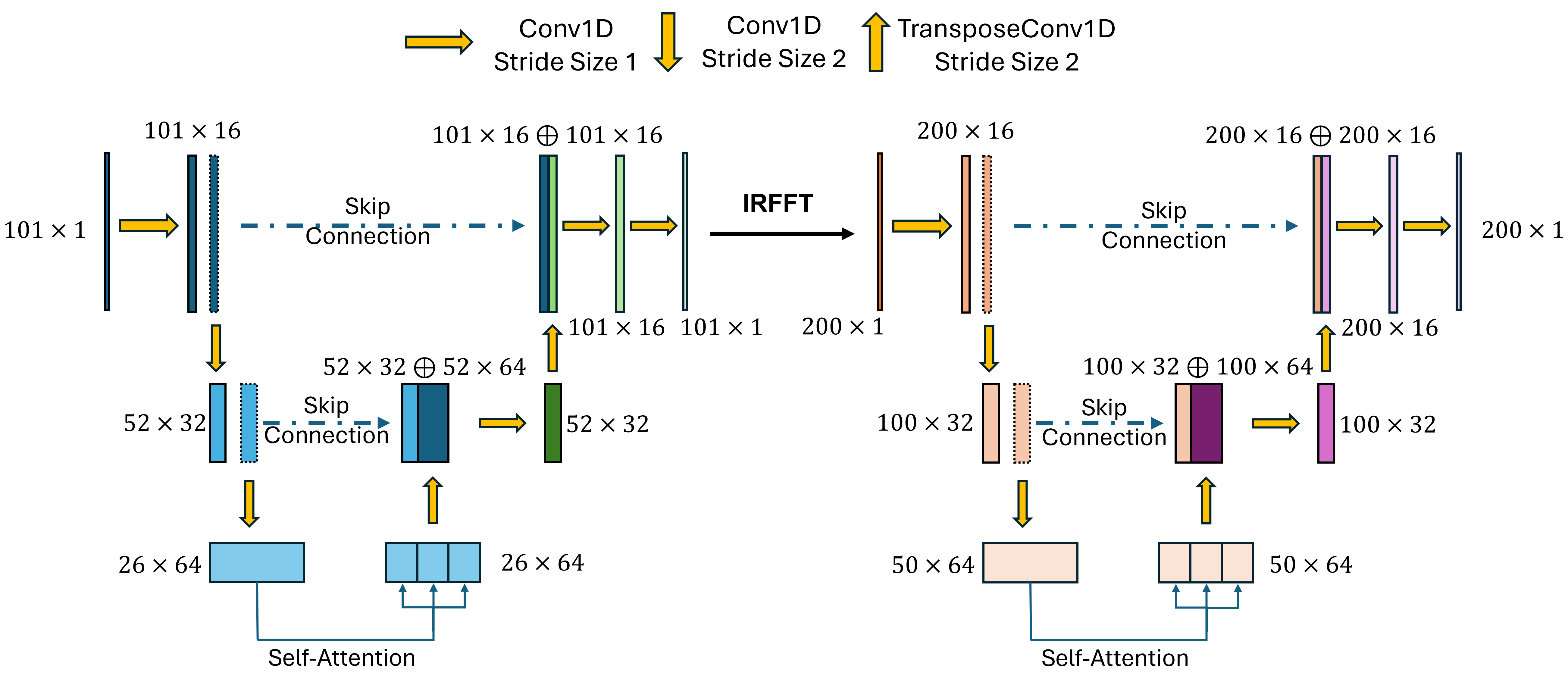}
\caption{\label{core}
\textbf{Core of the denoising network.} Stage~1 denoises the Fourier-domain representation and returns to the time domain. Stage~2 refines in the time domain using a UNet with a self-attention bottleneck to preserve long-range phase coherence and weak beating.}
\end{figure*}

The denoising network reconstructs the latent Ramsey waveform from noisy measurements while preserving the weak beating features that encode hyperfine-induced detunings. To address this, we use the two-stage architecture shown in Fig.~\ref{core}, in which an initial frequency-domain denoising step is followed by a time-domain refinement step.

\paragraph*{Two-stage denoising: frequency-domain coarse pass and time-domain refinement.}
For each noisy Ramsey trace, the first stage applies a real Fourier transform and performs a coarse denoising pass in the spectral domain using a compact 1D CNN--UNet. Operating in frequency space allows the model to attenuate broadband noise while retaining the narrowband beating structure associated with the underlying spin dynamics. The resulting coarse reconstruction is then transformed back to the time domain and passed to a second refinement network, which suppresses residual local artifacts and restores the correct oscillatory envelope. The details of the model structure is shown in Appendix~\ref{app:denoise_details}.

\paragraph*{Self-attention at the UNet bottlenecks.}
To complement convolutional locality with \emph{global} context, we augment the bottleneck of \emph{both} UNets with multi-head self-attention \cite{vaswani2017attention,Wang2018NonLocal,Bello2019AttentionAugmentedConv} within the standard encoder--decoder design \cite{ronneberger2015uUNet}. Functionally, this attention module allows the network to relate distant time points (or frequency bins) when forming its denoised representation, rather than relying purely on finite receptive fields \cite{Wang2018NonLocal,vaswani2017attention,Bello2019AttentionAugmentedConv}. This is particularly important for Ramsey traces, where the hyperfine-induced beating is expressed through distributed phase relationships and weak oscillatory components that can span the entire acquisition window. The full mathematical definition of the attention operator and our implementation details are given in Appendix~\ref{app:unet_attention}.

\subsubsection{\textbf{Training objective and optimization}}
\textbf{Normalization.} During training, each noisy/clean pair is min--max normalized using the minimum and maximum of the corresponding clean trace. This removes trivial amplitude scaling differences across synthetic configurations and focuses learning on waveform structure.

\textbf{Loss function.} The core is pretrained on the synthetic dataset (Sec.~\ref{sec:noisemodeling}) using paired noisy and clean traces. We minimize a time-domain mean-squared error together with a short-time Fourier transform (STFT) term that encourages preservation of local spectral content \cite{Yamamoto2020ParallelWaveGAN}:
\begin{equation}
\begin{aligned}
\mathcal{L}_{\mathrm{core}}
&=
\|\widehat{\mathbf{y}}-\mathbf{y}\|_2^2 
\quad\\ &+\;
\lambda_{\mathrm{STFT}}
\left\|
\mathrm{STFT}\!\left(\widehat{\mathbf{y}}-\langle\widehat{\mathbf{y}}\rangle\right)
-
\mathrm{STFT}\!\left(\mathbf{y}-\langle\mathbf{y}\rangle\right)
\right\|_{1}
\end{aligned}
\label{eq:loss_core}
\end{equation}
where $\widehat{\mathbf{y}}$ is the denoised output and $\mathbf{y}$ is the corresponding clean target. The first term is a \emph{mean-squared error} (MSE) in the time domain, which penalizes pointwise deviations between the predicted and clean traces. The second term is a \emph{short-time Fourier transform} (STFT) consistency penalty: After mean-centering each trace to remove trivial DC offsets (the DC component corresponds to the signal mean), we compute the STFT of the prediction and target and minimize their elementwise $\ell_1$ difference to encourage agreement of local time--frequency structure \cite{CambridgeDSP_DCmean,Engel2020DDSP,Yamamoto2020ParallelWaveGAN}. We set the weighting factor to $\lambda_{\mathrm{STFT}}=0.5$ in all experiments reported here.

\subsubsection{\textbf{Uncertainty adapters for simulation-to-reality fine-tuning}}
\begin{figure}[b]
\includegraphics[width=\linewidth]{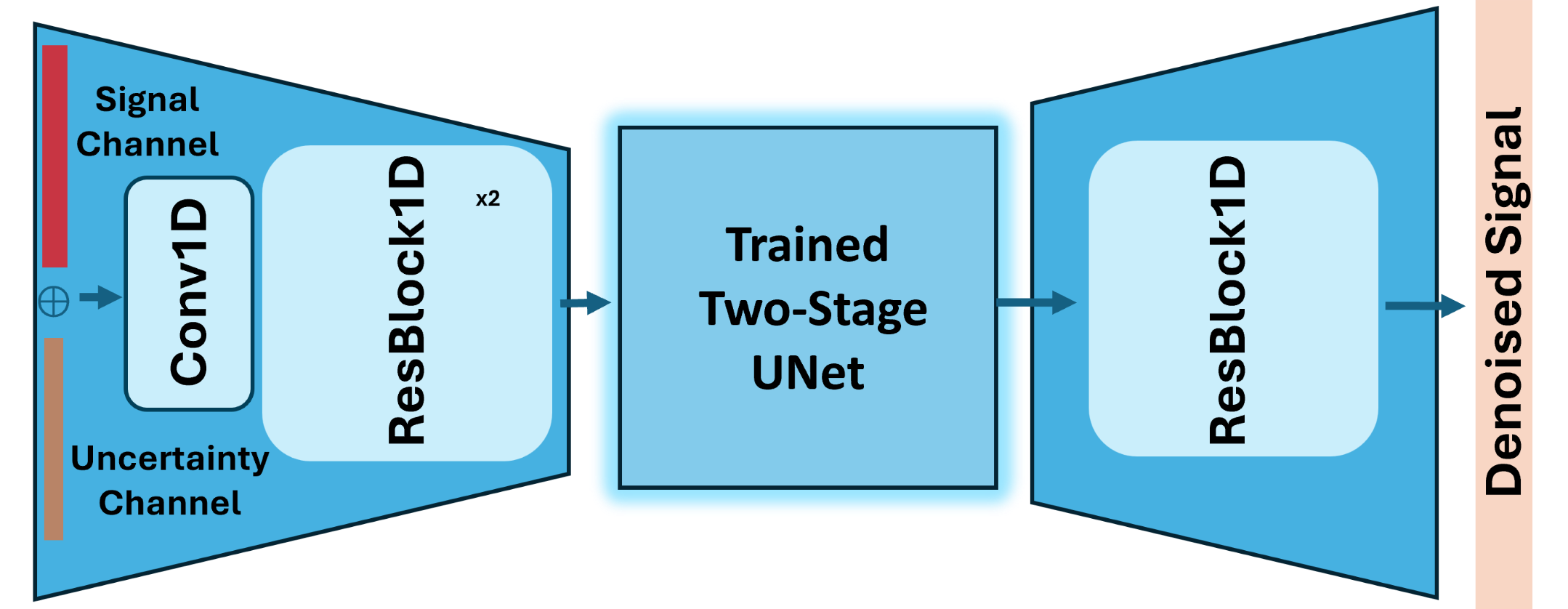}
\caption{\label{adapters}
\textbf{Adapters for experimental fine-tuning.} A front adapter ingests the noisy PL trace and an uncertainty channel and produces the core input. The core, composed of a pretrained UNet, is frozen during fine-tuning. A back adapter maps the output to the experimental trace format.}
\end{figure}

To bridge the residual domain gap between synthetic and experimental noise---including correlated drift, contrast fluctuations, and uncertainty miscalibration not fully captured by the analytic noise model---we adopted lightweight \emph{uncertainty adapters} that are fine-tuned on experimental traces while freezing the simulation-pretrained core network (Fig.~\ref{adapters}). Concretely, a \emph{front adapter} ingests the raw experimental PL trace together with a per-time-point uncertainty channel and maps this two-channel input into the feature space expected by the core. A \emph{back adapter} then maps the core output back into the experimental trace representation. This design concentrates the simulation-to-reality correction into a small number of trainable parameters, enabling stable adaptation without catastrophic forgetting of the physics-informed structure learned during large-scale simulation pretraining.

\subsubsection{\textbf{Denoising Result}}

\begin{figure*}[t!]
\includegraphics[width=6 in]{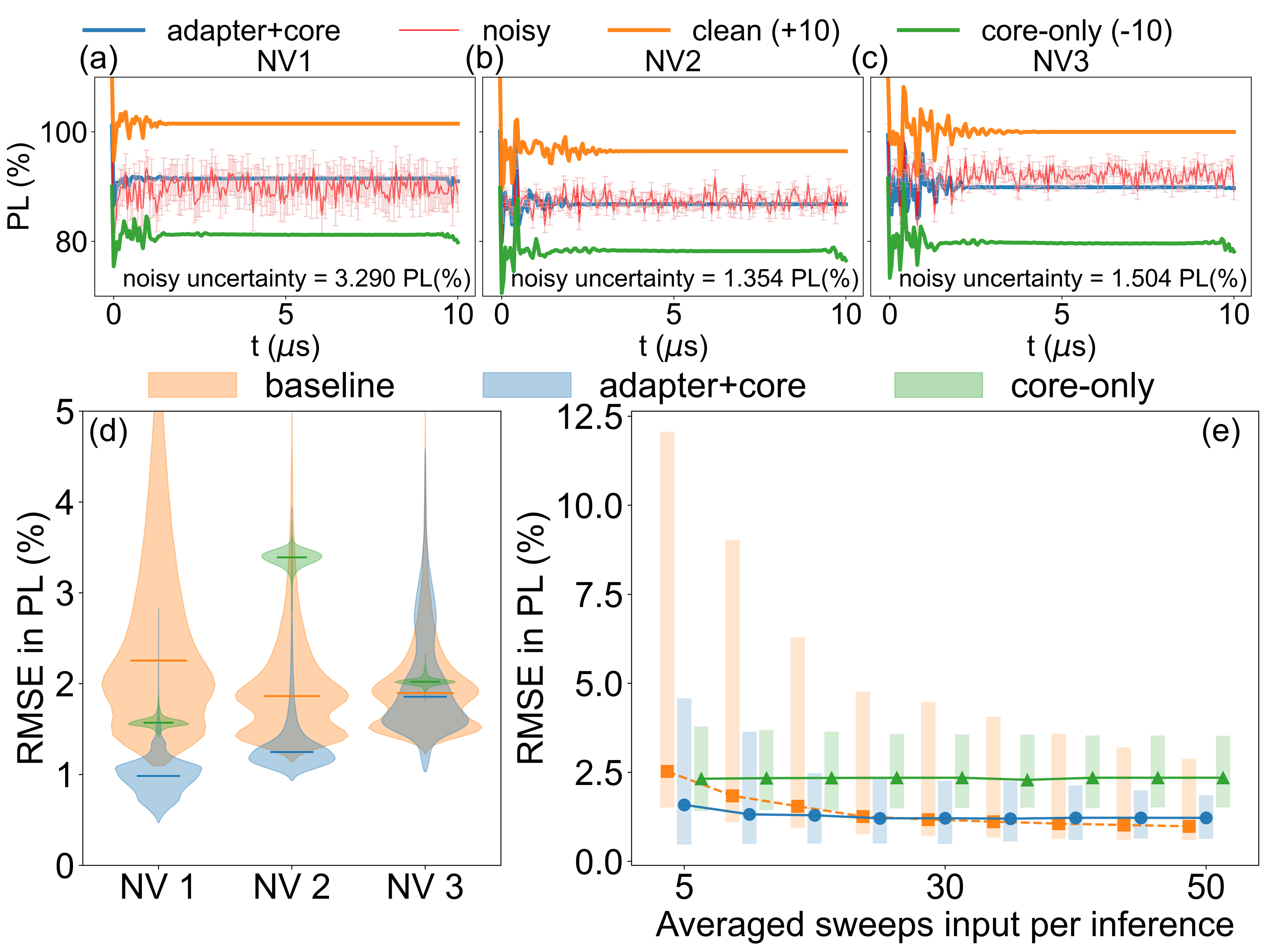}
    \caption{\label{denores}\textbf{Experimental denoising via simulation pretraining and adapter fine-tuning.}(a--c) Example Ramsey PL(\%) traces from three NV centers. Black: raw experimental input. Orange: 200-sweep fitted reference ($+10$ PL(\%) offset). Green: simulation-trained core-only output ($-10$ PL(\%) offset). Blue: core+adapter output after experimental fine-tuning. Offsets are for visualization only. (d) RMSE distributions (in PL percentage points) over all held-out few sweep test traces ($K\le 10$) for each NV, comparing the raw experimental baseline (orange), the simulation-trained core-only model (green), and the experimentally fine-tuned core+adapter model (blue). (e) RMSE versus number of averaged sweeps $K$ per inference.}
\end{figure*}

We evaluate denoising on experimental Ramsey PL time traces using two model variants: a \emph{core-only} network trained exclusively on simulated data with artificially added Gaussian noise, and a \emph{core+adapter} model in which the same simulation-pretrained core is augmented with lightweight adapter modules that are fine-tuned on experimental training traces. The experimental test set is strictly held out from all data used for adapter fine-tuning. This comparison directly tests the central hypothesis of this work: modest experimental adaptation on top of simulation pretraining produces a denoiser that is more robust to laboratory noise processes than a simulation-only model trained with Gaussian noise augmentation.

Denoising performance is quantified using the root-mean-square error (RMSE) between the denoised trace $\hat{y}(t)$ and a high-SNR reference trace $y(t)$, reported in percentage points of normalized PL. For discretely sampled evolution times $\{t_i\}_{i=1}^{T}$, we compute
\begin{equation}
\mathrm{RMSE}(\hat{y},y)=\sqrt{\frac{1}{T}\sum_{i=1}^{T}\left(\hat{y}(t_i)-y(t_i)\right)^2},
\end{equation}
which summarizes the typical pointwise deviation of the denoised waveform from the best available experimental estimate of the underlying Ramsey response. As the reference $y(t)$, we use the least-squares fit to a dedicated high-averaging acquisition for each measurement (orange curve in Fig.~\ref{denores}), which provides a consistent high-SNR target across datasets. Lower RMSE, therefore, corresponds to smaller residual mismatch to the experimental reference trace and, in practice, to improved stability for downstream fitting of quantities such as $T_2^\ast$, detuning, and hyperfine beating components. Further discussion of the relation between this metric and residual-based fitting diagnostics is provided in Appendix~\ref{app:rmse_chi2}.

Figure~\ref{denores}(a)--(c) shows representative single-trace examples from three NV centers. For visual clarity only, the reference and the core-only output are plotted with a constant $+10\%$ vertical offset; this offset is not applied when computing RMSE. In each case, the raw experimental input (black) exhibits substantial point-to-point fluctuations that obscure the Ramsey envelope. The core+adapter output (blue) suppresses these fluctuations while preserving the qualitative features of the trace, yielding a lower RMSE than the raw baseline in all three examples: for NV~1 the RMSE decreases from $2.560$ to $0.470$~PL(\%), for NV~2 from $1.629$ to $0.936$~PL(\%), and for NV~3 from $2.474$ to $1.031$~PL(\%). In contrast, the simulation-only core model (green dashed) is consistently worse than the core+adapter model on these experimental examples, most notably for NV~2 where it incurs a large error ($3.400$~PL(\%) ). These qualitative examples already indicate that the dominant failure mode is not the network capacity per se, but rather the simulation-to-experiment domain gap: Gaussian noise injection captures part of the stochastic variance but does not reproduce common experimental nuisances such as correlated readout fluctuations, slow drift, contrast variations, and other device-specific artifacts. Adapter fine-tuning provides a lightweight mechanism to absorb these non-idealities without retraining the full model.

The dataset-level behavior in the few sweep setting (inputs formed by averaging only $K\le 10$ experimental sweeps) is summarized in Fig.~\ref{denores}(d), which compares RMSE distributions over all held-out traces for each NV. In this noise-dominated regime, the core+adapter model improves upon the experimental baseline for all three NVs, indicating that adapter fine-tuning learns experimentally relevant noise statistics that are not captured by simulation-only pretraining.  We refer to the ``experimental baseline" as the no-denoising reference in which the raw experimental input trace is directly compared to the clean reference $\mathrm{PL}_{\mathrm{ref}}(t)$ when computing RMSE.
Qualitatively, Table~\ref{tab:denoise_rmse_nv} shows that the adapter-fine-tuned model consistently reduces denoising error relative to the raw experimental baseline for all three NV centers.
The improvement is most pronounced for NV~1 and NV~2, where the error distributions shift substantially toward lower RMSE after adaptation, indicating that the adapters effectively absorb NV-specific experimental noise processes. For NV~3 the gain is more modest, but the median behavior still favors the adapted model, suggesting that the core already captures a larger fraction of the dominant noise structure for this dataset and that only limited additional correction is required.
This few-sweep regime is operationally central for rapid NV characterization, where parameter inference must be performed with minimal acquisition overhead; reducing typical denoising error therefore directly stabilizes downstream fits (e.g., $T_2^*$ and detuning) and suppresses rare large-error excursions that can otherwise dominate least-squares estimates.

By contrast, the simulation-trained core-only model is not a reliable deployment strategy under $K\le 10$ conditions. While it improves relative to baseline for NV~1, it fails markedly for NV~2 (median $3.392$~PL(\%)), and it remains worse than baseline for NV~3. Beyond median performance, Fig.~\ref{denores}(d) highlights a qualitative limitation of simulation-only training: the core-only RMSE distributions are comparatively narrow and weakly dependent on the NV, suggesting that the model does not strongly \emph{respond} to changes in experimental noise conditions. In effect, a network trained only on simulated noise tends to produce outputs with a more uniform error profile even when the true experimental noisiness varies across NVs and traces. Adapter fine-tuning mitigates this domain mismatch by incorporating experimental noise statistics into the mapping, yielding both lower typical error in the few sweep regime and an error distribution that more faithfully reflects the heterogeneity of real measurements.

Figure~\ref{denores}(e) probes robustness as a function of measurement SNR by varying the number of averaged experimental sweeps $K$ used to form each input trace prior to inference. From an experimental perspective, this analysis is operationally important: it delineates the regime in which the ML denoiser provides a genuine reduction in acquisition overhead (i.e., achieving a target reconstruction error with substantially fewer sweeps) versus the regime in which conventional averaging already yields high-SNR traces and the denoiser offers little additional benefit. Consistent with statistical averaging, the experimental baseline error decreases monotonically with $K$. The core+adapter model exhibits its strongest advantage in the low-$K$ (noise-dominated) regime that is most relevant for rapid characterization, and it continues to improve as $K$ increases, indicating that the learned mapping does not saturate at a fixed denoising strength but can exploit higher-quality inputs. This behavior is especially significant because the adapters are conditioned on a per-time-point uncertainty channel: as the effective uncertainty decreases with increasing $K$, the model can adaptively reduce the degree of correction, whereas at higher uncertainty it applies a stronger suppression of stochastic fluctuations while preserving coherent beating structure. In contrast, the simulation-only core model shows only weak dependence on $K$, suggesting limited sensitivity to the true experimental noise level and, correspondingly, limited ability to capitalize on additional averaging. At sufficiently large $K$, where stochastic noise is already strongly suppressed, the residual discrepancy can become dominated by small systematic effects (e.g., mild over-smoothing or a subtle amplitude/offset bias), so the marginal advantage of the core+adapter model may diminish and can even slightly reverse. Taken together, panels (d) and (e) identify the practical ``assist'' window for ML-based denoising and highlight that uncertainty-conditioned adapter fine-tuning restores a physically reasonable dependence of reconstruction performance on measurement SNR, which is largely absent under simulation-only training with Gaussian augmentation.

\begin{table}[t]
\caption{\label{tab:denoise_rmse_nv}\textbf{few sweep denoising performance on Independent test experimental traces.}
Medians of per-trace RMSE (in PL percentage points) are reported for inputs constructed from $\leq 10$ averaged sweeps ($K\leq 10$), with sample counts $n$.}
\begin{ruledtabular}
\setlength{\tabcolsep}{3.5pt}
\renewcommand{\arraystretch}{1.05}
\scriptsize
\begin{tabular}{lcccc}
NV & $n$ & Base med. & Core med. & Core+Adapt med. \\
\hline
NV~1 & 40000 & 2.255 & 1.570 & \textbf{0.984} \\
NV~2 & 40000 & 1.864 & 3.392 & \textbf{1.248} \\
NV~3 & 40000 & 1.897 & 2.021 & \textbf{1.856} \\
\end{tabular}
\end{ruledtabular}
\end{table}

\subsection{\label{sec:level2}Hyperfine Parameter Estimator}
\label{sec:hf_estimator}
\subsubsection{\textbf{Model structure and training}}

\begin{figure*}[t!]
\includegraphics[width=6 in]{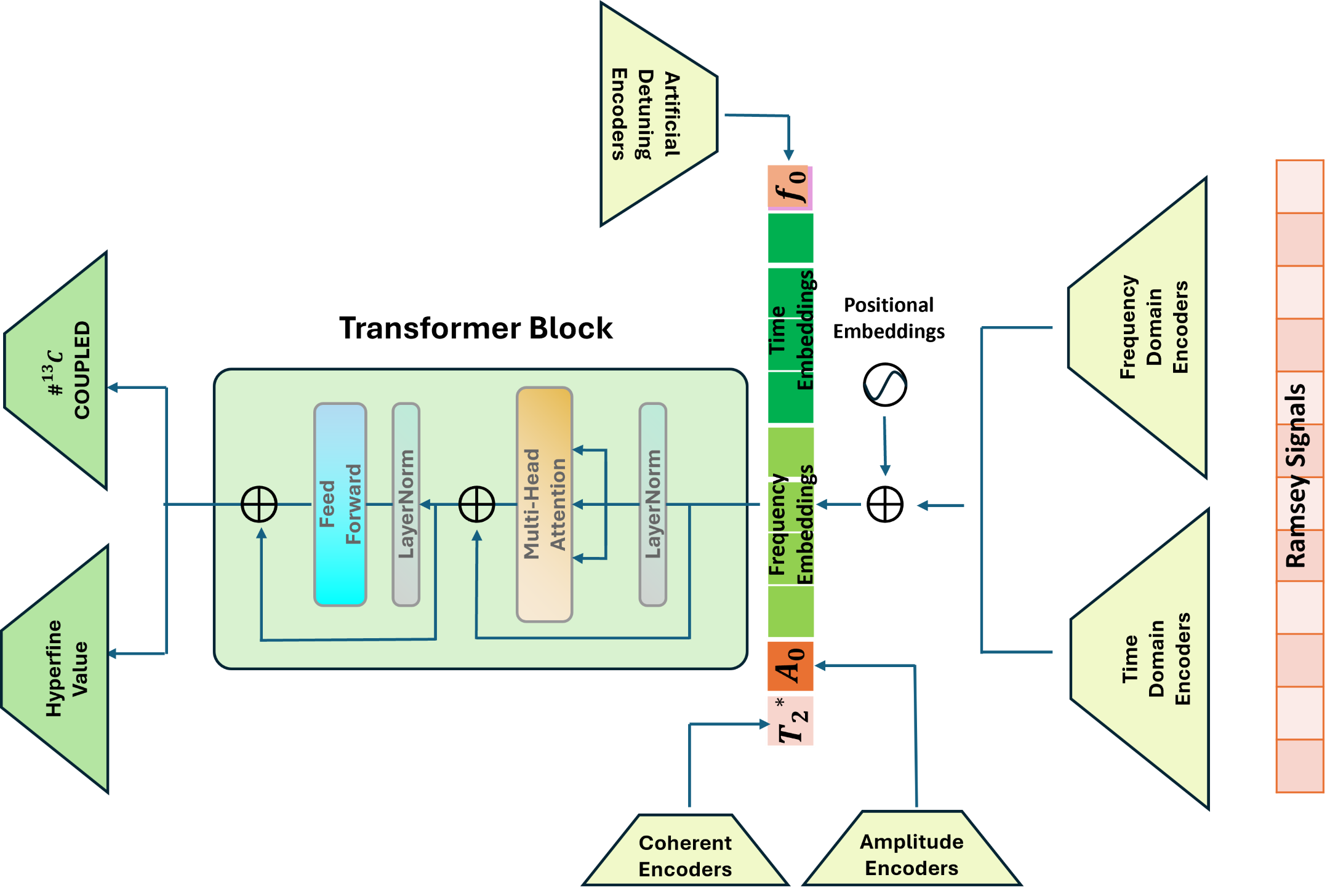}
\caption{\label{HFPredictionStruc}
\textbf{Architecture of the hyperfine-parameter estimator.}
A denoised Ramsey trace is processed by three auxiliary ``oscillation-pattern'' encoders that regress global waveform factors: the inhomogeneous dephasing time $T_2^\ast$, the overall PL/contrast amplitude, and a small offset detuning $f$ that accounts for experimental drift. These scalars are embedded as metadata tokens and concatenated with time-domain tokens (200 normalized samples) and frequency-domain tokens (101 log-magnitude rFFT bins) to form a 304-token sequence. A Transformer encoder then performs joint time--frequency self-attention and outputs a pooled representation, from which two heads predict (i) the discrete ${}^{13}\mathrm{C}$ count and (ii) a fixed-length, zero-padded vector of parallel hyperfine couplings.}

\end{figure*}

After denoising, the remaining task is to infer the hyperfine content that produces the observed beating pattern in Ramsey signals. We cast this as a supervised parameter-estimation problem trained entirely on simulation data, where the ground-truth hyperfine configuration is known. The estimator is modular (Fig.~\ref{HFPredictionStruc}): three ``oscillation pattern encoders'' extract scalars that set the gross shape of the trace (overall contrast level, inhomogeneous dephasing time, and an experimental offset detuning), and a final Transformer-based predictor converts these summaries together with the waveform features into hyperfine parameters.

\paragraph*{Oscillation pattern encoders (auxiliary regressors).}
These encoders provide physically interpretable scalars that strongly affect the Ramsey waveform, and they are trained on simulation metadata.

\textbf{(i) Inhomogeneous dephasing time encoder ($T_2^*$).}
We use a one-dimensional convolutional neural network (1D CNN) followed by a Transformer encoder layer and a small multilayer perceptron (MLP). A 1D CNN applies the same short filter along the time axis, so it is well matched to local oscillatory features while remaining insensitive to absolute time shifts. In our implementation each convolution stage is followed by a BatchNorm layer\cite{IoffeSzegedy2015BatchNorm} and Gaussian Error Linear Unit (GELU) \cite{hendrycks2016gelu}.
 A single Transformer encoder layer then aggregates information over the entire trace before mean pooling and regression to $T_2^*$.

The $T_2^*$ model is trained using a mean-squared error in \emph{log space},
\begin{equation}
\mathcal{L}_{T_2^*}
=
\left\|\log(1+\widehat{T}_2^*)-\log(1+T_2^*)\right\|_2^2,
\end{equation}
which reduces sensitivity to the wide dynamic range of $T_2^*$ values and makes relative errors more uniform across the dataset\cite{BoxCox1964}.

\textbf{(ii) Overall amplitude (contrast) encoder.}
The PL/contrast level is inferred by a compact 1D CNN with global average pooling and an MLP head. The network concatenates a normalized time channel (time rescaled to $[0,1]$ per trace) so it can learn consistent temporal priors. This encoder is trained with standard mean-squared error:
\begin{equation}
\mathcal{L}_{\mathrm{PL}}=\|\widehat{\mathrm{PL}}-\mathrm{PL}\|_2^2.
\end{equation}

\textbf{(iii) Experimental offset detuning encoder ($f$ estimator).}
In experiments we apply an artificial detuning near $5$~MHz to simplify Fourier-domain analysis (see Appendix.~\ref{DatAug} and Fig.~\ref{fig:ramsey_sequence_appendix}); however, microwave drift and timing imperfections can shift the apparent central peak by up to $\pm 0.5$~MHz. We explicitly model this by training a dedicated estimator on simulated data where the detuning is randomized within this range. The model computes an internal real Fourier transform, converts the spectrum to a log-magnitude representation (to compress dynamic range), and then uses a small 1D CNN and an MLP to regress the offset frequency $\widehat{f}$ (in MHz). We train with a robust regression loss (SmoothL1):
\begin{equation}
\mathcal{L}_{f}=\mathrm{SmoothL1}(\widehat{f},f).
\end{equation}
Defining the residual $r=\widehat{f}-f$, the SmoothL1 penalty is
\begin{equation}
\mathrm{SmoothL1}(r)=
\begin{cases}
\frac{1}{2}\,r^2/\beta, & |r|<\beta,\\[4pt]
|r|-\frac{1}{2}\beta, & |r|\ge \beta,
\end{cases}
\end{equation}
where $\beta>0$ sets the transition between quadratic and linear behavior. Thus the loss is locally quadratic for small errors (encouraging precise fits) and becomes linear for large errors (reducing sensitivity to outliers) \cite{liu2016ssdl1,girshick2015fastl1,huber1992robustsl1}. In our implementation we use the default $\beta=1$ (in MHz units).

\paragraph*{Hyperfine frequency predictor (Transformer head).}
As shown in Fig.~\ref{HFPredictionStruc}, the hyperfine frequency predictor maps a denoised Ramsey trace to (i) the discrete number of coupled $^{13}$C spins $n\in\{0,\dots,10\}$ and (ii) a fixed-length vector of hyperfine parameters $\widehat{\mathbf{c}}\in\mathbb{R}^{10}$ (zero-padded when fewer spins are present). We evaluate traces in mini-batches; here $\text{batchsize}$ denotes the number of traces processed in parallel.

\paragraph*{Tokenization and Transformer predictor.}
The Transformer head operates on a joint time--frequency representation of each \emph{denoised} Ramsey trace, augmented with the physically interpretable scalars produced by the auxiliary encoders (Fig.~\ref{HFPredictionStruc}). Concretely, we first normalize each trace to remove trivial amplitude and offset variation, then represent it using two complementary views: (i) the normalized time-domain samples, which preserve phase evolution and envelope shape, and (ii) a compact frequency-domain summary derived from the real Fourier transform, which highlights narrowband beating components associated with hyperfine-induced detunings. We concatenate these waveform tokens with a learnable classification summary token and a small set of metadata tokens encoding the estimated global factors (contrast/PL, $T_2^\ast$, and the offset detuning $f$) \cite{Devlin2019BERT, Dosovitskiy2021ViT}. A Transformer encoder \cite{vaswani2017attention} then performs self-attention over the full token sequence, enabling the model to relate distributed phase information in the time domain to localized spectral structure in the frequency domain. The final classification embedding is used as a pooled sequence representation \cite{Devlin2019BERT,Dosovitskiy2021ViT,Touvron2021DeiT} and is passed to two heads that predict (i) the discrete ${}^{13}$C count and (ii) a fixed-length, zero-padded vector of parallel hyperfine couplings. The precise token construction, embedding maps, and the attention equations are provided in Appendix~\ref{app:hf_tokens}.

\textbf{Count head and cross-entropy.}
The count head treats the ${}^{13}\mathrm{C}$ number as a discrete label \(n\in\{0,\dots,10\}\). 
For each trace in a mini-batch, it outputs an 11-element score vector (logits) 
\(\widehat{\mathbf{z}}^{(b)}\in\mathbb{R}^{11}\), where \(b\in\{1,\dots,\text{batchsize}\}\) indexes the
\(b\)-th sample (trace) in the mini-batch. We convert these scores into class probabilities with a softmax \cite{bridle1989trainingsoftmax},
\begin{equation}
p^{(b)}_j=\frac{\exp\!\big(\widehat{z}^{(b)}_j\big)}{\sum_{k=0}^{10}\exp\!\big(\widehat{z}^{(b)}_k\big)},
\qquad j\in\{0,\dots,10\},
\end{equation}
and train the head using the cross-entropy loss \cite{ackley1985learningcorssentropy,hopfield1987learningcrossentropy,baum1987supervisedcorssentropy,levin1988accelerated},
\begin{equation}
\mathcal{L}_{\mathrm{count}}
=
-\frac{1}{\text{batchsize}}\sum_{b=1}^{\text{batchsize}}
\log p^{(b)}_{\,n^{(b)}},
\label{eq:ce_def}
\end{equation}
where \(n^{(b)}\in\{0,\dots,10\}\) is the ground-truth ${}^{13}\mathrm{C}$ count label for the \(b\)-th trace.

Physically, this objective encourages the model to place high probability on the correct nuclear-spin
count class for each denoised Ramsey trace. Different ${}^{13}\mathrm{C}$ counts correspond to different
levels of beating complexity in the Ramsey free-induction signal: low-count environments typically yield
sparser modulation components, while higher counts produce denser multi-frequency structure that is more
ambiguous in few-shot data. Cross-entropy penalizes the network when it assigns low probability to the
true count, promoting discriminative features that separate these regimes while allowing uncertainty to be
expressed through probability mass distributed across neighboring count classes when the waveform is
intrinsically ambiguous.

\textbf{Hyperfine head and masked regression.}
The hyperfine head outputs \(\widehat{\mathbf{c}}\in\mathbb{R}^{\text{batchsize}\times 10}\) (kHz). Because the target vector is zero-padded when fewer spins are present, we apply a masked regression loss. For a sample with true count \(n\), define a mask
\begin{equation}
m_i=
\begin{cases}
1, & i\le n,\\
0, & i>n,
\end{cases}
\qquad i=1,\dots,10,
\end{equation}
and use a masked SmoothL1 objective\cite{liu2016ssdl1,huber1992robustsl1,girshick2015fastl1}
\begin{equation}
\mathcal{L}_{\mathrm{cij}}
=\frac{1}{\sum_{i=1}^{10} m_i}\sum_{i=1}^{10} m_i\;\mathrm{SmoothL1}\!\left(\widehat{c}_i-c_i\right).
\end{equation}

\textbf{Total loss.}
The total multi-task loss is
\begin{equation}
\mathcal{L}_{\mathrm{HF}}
=
w_{\mathrm{count}}\;\mathcal{L}_{\mathrm{CE}}(\widehat{\mathbf{z}},\mathbf{n})
+
w_{\mathrm{cij}}\;\mathcal{L}_{\mathrm{cij}},
\label{eq:total_loss}
\end{equation}
so that (i) the count head learns a categorical distribution over \(n\in\{0,\dots,10\}\) via \(\mathcal{L}_{\mathrm{CE}}\) (Eq.~\ref{eq:ce_def}), and (ii) only the first \(n\) hyperfine entries contribute to \(\mathcal{L}_{\mathrm{cij}}\) for a sample with \(n\) coupled spins.

\subsubsection{\textbf{Parameter Estimation Results}}
\begin{figure}[t]
\includegraphics[width=0.5\textwidth,]{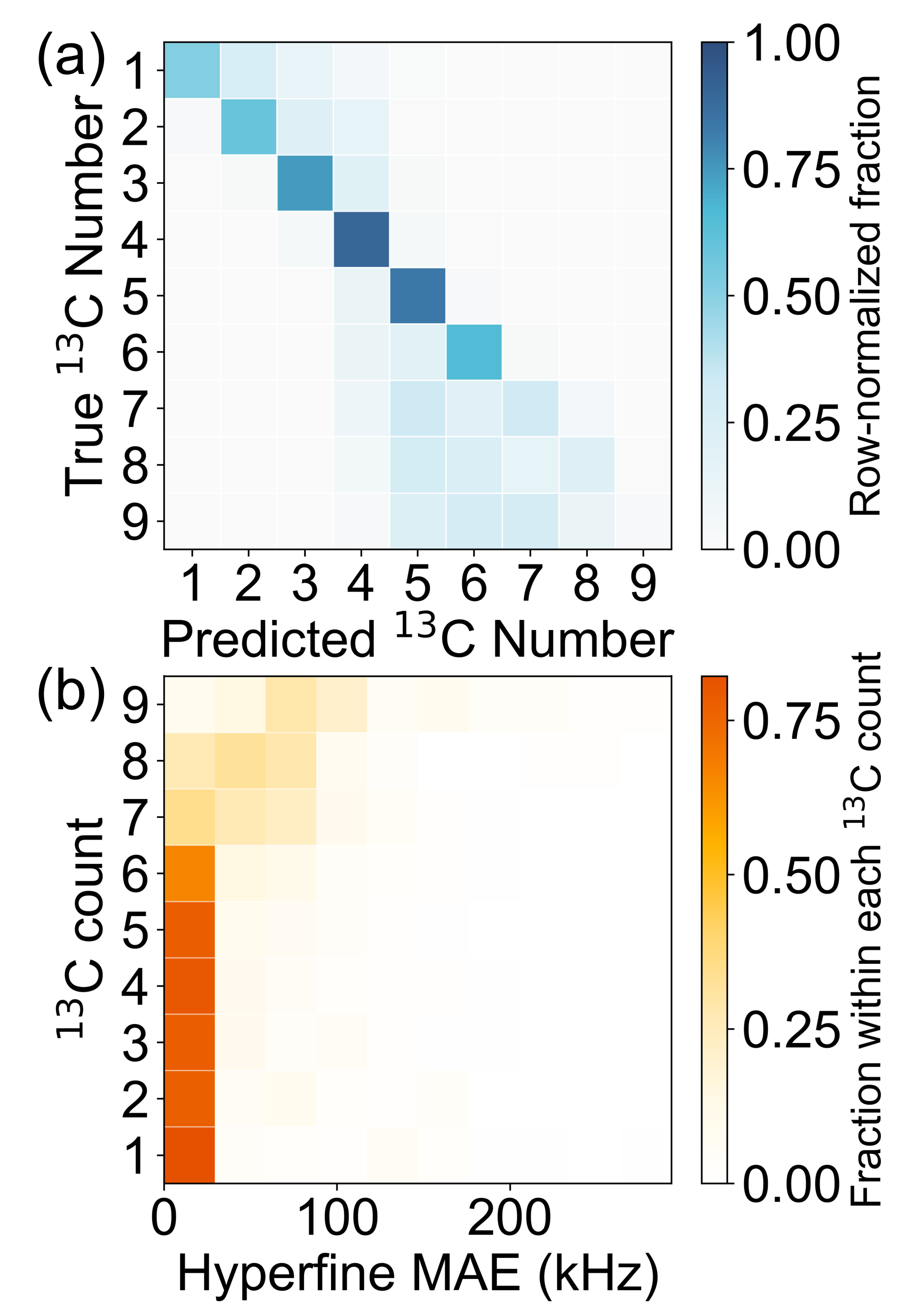}
\caption{\label{sim-C13PRE}\textbf{Hyperfine prediction on a held-out simulation test set.}
Test set: $135{,}000\times 3$ Ramsey traces spanning $\sim 450$ distinct ${}^{13}\mathrm{C}$ configurations ($n_{\max}=9$).
(a) Confusion matrix for ${}^{13}\mathrm{C}$ count $n$; intermediate counts are classified with high accuracy (e.g., $90.8\%$ at $n=4$), while large-$n$ cases show increased ambiguity (e.g., $39.4\%$ at $n=7$, $28.1\%$ at $n=8$).
(b) Distribution of hyperfine MAE for the parallel couplings $A_{\parallel}$ (kHz) versus true $n$, showing a broadened error tail as $n$ increases.}
\end{figure}

\textbf{Simulation results}
We first quantify hyperfine-parameter inference on a fully independent simulated noise-free test set. The evaluation comprises $\sim 500,000$ Ramsey traces, spanning $\sim 450$ distinct ${}^{13}\mathrm{C}$ doping configurations with a maximum count of $n_{\max}=9$. The model produces (i) a discrete estimate of the ${}^{13}\mathrm{C}$ number $n\in\{1,\dots,9\}$ and (ii) a continuous estimate of the parallel hyperfine couplings $\mathbf{A}_{\parallel}=(A_{\parallel,1},\dots,A_{\parallel,n})$, reported in kHz. For regression we use the per-trace mean absolute error (MAE)\cite{rumelhart1986learningMAE},
\begin{equation}
\mathrm{MAE}(A_{\parallel})=\frac{1}{n}\sum_{j=1}^{n}\left|A_{\parallel,j}-\hat{A}_{\parallel,j}\right|,
\end{equation}
where the sum runs over the $n$ nuclei in the ground-truth configuration and hats denote predictions.

Figure~\ref{sim-C13PRE}(a) reports the confusion matrix for ${}^{13}\mathrm{C}$-count prediction on this held-out test set. The network achieves its best performance at intermediate counts, with strong diagonal entries (e.g., $76.0\%$ for $n=3$, $90.8\%$ for $n=4$, $85.5\%$ for $n=5$, and $71.0\%$ for $n=6$). Misclassifications are predominantly near-diagonal, indicating that the dominant failure mode is confusion between neighboring counts rather than arbitrary errors. In contrast, performance degrades toward the distribution edges: low counts are frequently overestimated (e.g., $n=1$ is often mapped to $n=2$), while higher counts exhibit systematic under-resolution, with notably reduced diagonal fractions for $n\ge 7$ (e.g., $39.4\%$ for $n=7$ and $28.1\%$ for $n=8$). This trend is consistent with the increased spectral congestion of the Ramsey beating pattern as additional nuclei contribute, which makes distinct configurations less identifiable. The continuous regression behavior is summarized in Fig.~\ref{sim-C13PRE}(b), which shows the distribution of hyperfine MAE versus the true ${}^{13}\mathrm{C}$ count. The MAE is concentrated at low values for $n\le 6$ and broadens substantially for larger $n$, consistent with the growing parameter degeneracy and reduced identifiability at high nuclear density. Taken together, panels (a) and (b) demonstrate that the model captures physically meaningful structure in the simulated Ramsey response: regimes with accurate count classification also exhibit tight hyperfine-error distributions, whereas high-$n$ regimes suffer both increased count ambiguity and heavier MAE tails.

\textbf{Experimental results}

\begin{figure*}[t]
\includegraphics[width=6 in]{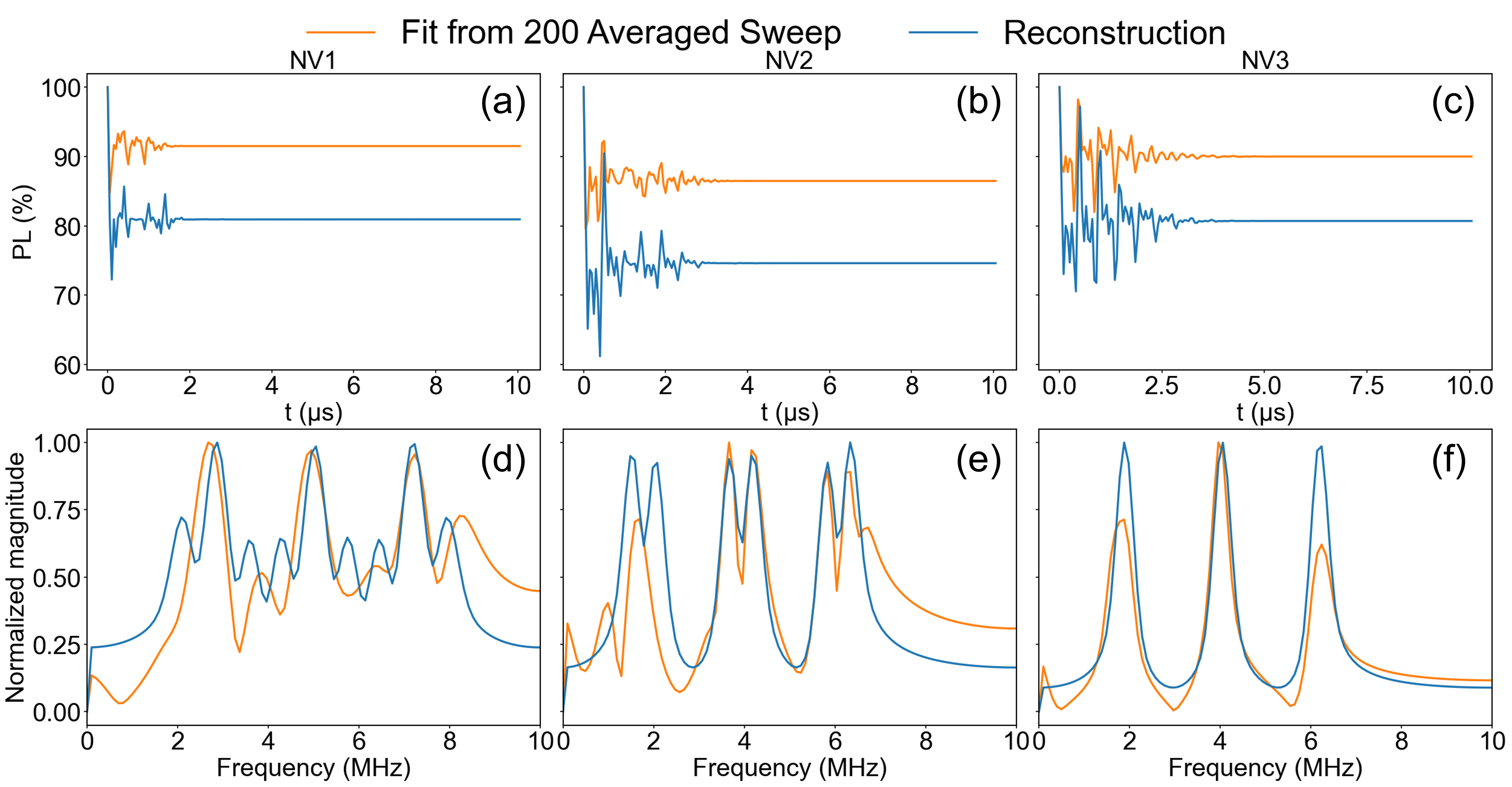}
\caption{\label{fig:exp_hf_recon}\textbf{Experimental validation of hyperfine-parameter estimation by forward reconstruction.}
(a--c) Time-domain Ramsey PL(\%) traces: fitted curve from 200 sweeps averages(orange) versus reconstruction(blue) from predicted hyperfine parameters for NV~1--3. 
(d--f) Corresponding normalized FFT magnitudes, showing agreement of the dominant hyperfine features. Offsets/normalization are for visualization only.}
\end{figure*}


\begin{table}[t]
\caption{\label{tab:exp_hf_params}\textbf{Representative experimental hyperfine-parameter estimates and $T_2^*$ comparison.}
For each NV we report a representative example, including the number of averaged sweeps used as input ($K$), the estimated $^{13}$C count $\hat{N}_{^{13}\mathrm{C}}$, and the inferred hyperfine-parallel couplings $\hat{C}_{ij}$. We also compare the experimentally fitted $T_2^*$ (with uncertainty) to the reconstruction-scan prediction $T_{2,\mathrm{pred}}^*$. The frequency-domain reconstruction error is quantified by $\mathrm{RMSE}_{\mathrm{FFT}}$ computed on the normalized FFT magnitude.}
\begin{ruledtabular}
\setlength{\tabcolsep}{3pt}
\renewcommand{\arraystretch}{1.15}
\scriptsize
\begin{tabular}{lcccccc}
NV 
& $K$ 
& $\hat{N}_{^{13}\mathrm{C}}$
& $\hat{C}_{ij}$
& $\mathrm{RMSE}_{\mathrm{FFT}}$
& $T_{2,\mathrm{pred}}^*$
& $T_{2,\mathrm{exp}}^*$ \\
& & & (MHz) & & ($\mu$s) & ($\mu$s) \\
\hline
NV~1 & 5 & 3 & $[0.79,\, 0.65,\, -0.02]$ & 0.18 & 1.13 & $0.91 \pm 0.169$ \\
NV~2 &  5 & 2 & $[0.05,\,-0.45]$ & 0.18 & 1.57 & $1.42 \pm 0.095$ \\
NV~3 &  5 & 1 & $[0.04]$          & 0.10 & 1.74  & $1.94 \pm 0.335$ \\
\end{tabular}
\end{ruledtabular}
\end{table}

The experimental performance of the hyperfine-parameter estimator is summarized in Fig.~\ref{fig:exp_hf_recon} and Table~\ref{tab:exp_hf_params}. For each NV center, we input a denoised Ramsey trace to the network to obtain a predicted ${}^{13}\mathrm{C}$ count $\hat{N}_{{}^{13}\mathrm{C}}$ and a set of parallel hyperfine couplings $\{\hat{C}_{ij}\}$. To validate these predictions, we perform a forward reconstruction: the inferred parameters are inserted into the same reduced NV Hamiltonian model used in simulation, generating a synthetic Ramsey response under the experimental timing grid. In Fig.~\ref{fig:exp_hf_recon}(a--c), this reconstructed curve is compared directly to the high-SNR experimental reference defined by a least-squares fit to the 200-sweep averaged measurement on the same NV (Sec.~\ref{sec:noisemodeling}). The close agreement in the time domain, both in the overall decay envelope and in the early-time oscillatory pattern, indicates that the inferred hyperfine parameters capture the dominant coherent components of the experimental signal, rather than simply producing a smoothed waveform.

The frequency-domain comparison provides a more direct diagnostic of hyperfine content: the normalized FFT magnitudes [Fig.~\ref{fig:exp_hf_recon}(d--f)] show substantial overlap of the principal hyperfine bands across all three NV centers, with small $\mathrm{RMSE}_{\mathrm{FFT}}$ values (Table~\ref{tab:exp_hf_params}) that quantitatively support the visual agreement. In particular, the dominant spectral features and their approximate locations are preserved, consistent with the estimator recovering physically meaningful hyperfine-parallel couplings that explain the observed modulation frequencies.

Operationally, this result is significant because hyperfine-assisted Ramsey fitting in real experiments often requires extensive manual initialization and iterative “guess-and-check” of discrete $^{13}$C configurations and coupling parameters to avoid local minima and unphysical solutions. The predictor provides a principled, data-driven initialization: it outputs a plausible $\hat{N}_{^{13}\mathrm{C}}$ and a ranked set of $\hat{C}_{ij}$ values that can be used directly as starting points (or priors) for conventional nonlinear least-squares refinement. Because the network is trained on a carefully constructed physics-based simulation set spanning hundreds of distinct $^{13}$C configurations, these estimates are constrained to lie within a physically consistent manifold, which helps regularize the inference problem and reduces the human overhead in parameter tuning. The representative examples in Table~\ref{tab:exp_hf_params} further illustrate that the model produces interpretable outputs at low averaging ($K=5$), including nontrivial multi-carbon solutions (e.g., $\hat{N}_{^{13}\mathrm{C}}=5$ for NV~1 and NV~2) and a lower-count case (NV~3), while still maintaining strong FFT agreement.

\subsubsection{\textbf{Evaluation on Auxiliary Encoders}}

\begin{figure*}[t]
\includegraphics[width=6 in]{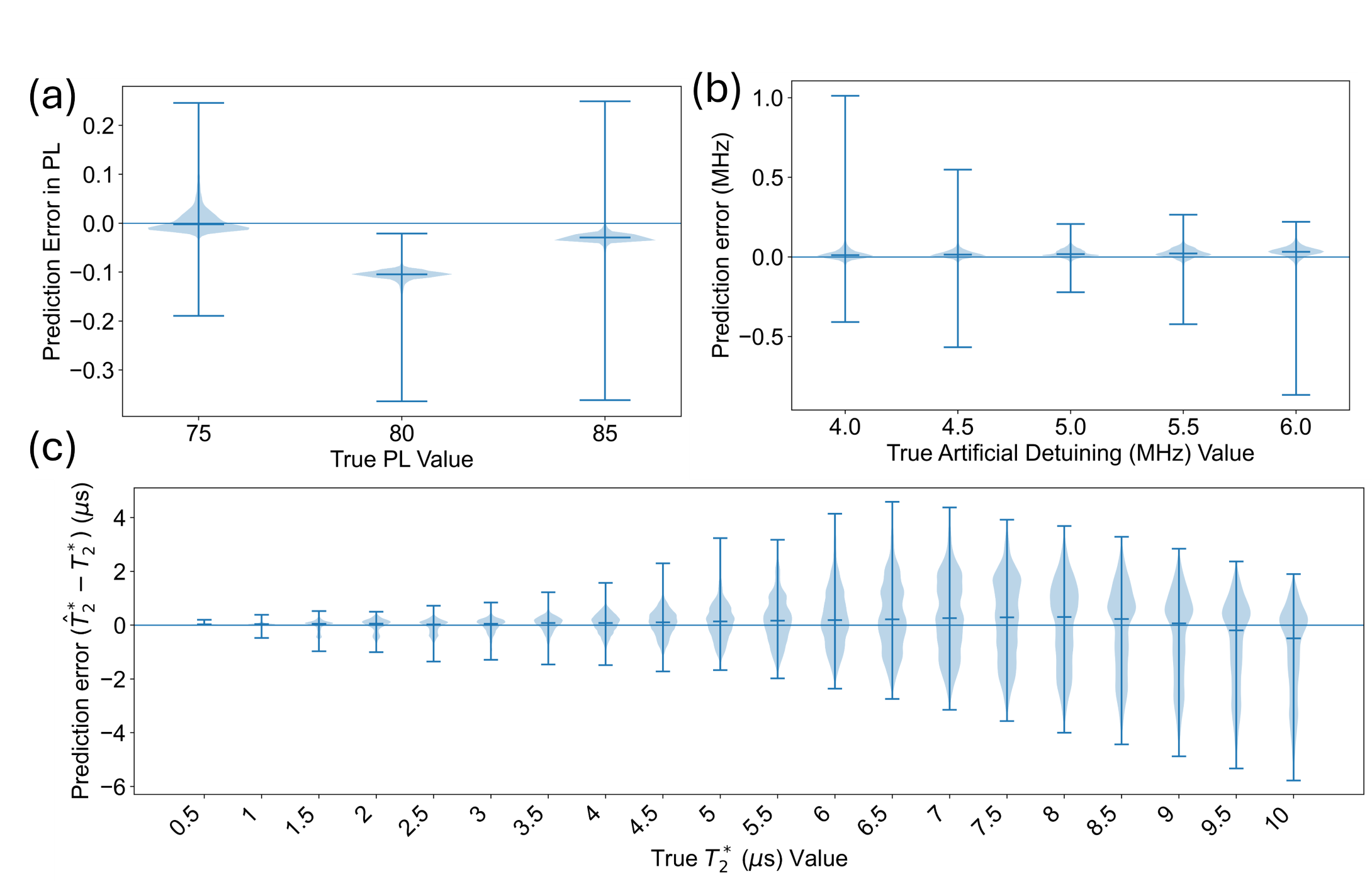}
\caption{\label{fig:auxiressim}
\textbf{Simulation performance of the three auxiliary ``oscillation-pattern'' encoders.}
Violin plots summarize prediction errors on a held-out simulated test set, reported as \emph{prediction minus ground truth} at discrete label values (horizontal axes). (a) Overall PL/contrast encoder: error in PL units evaluated at representative PL operating points (75--85PL(\%)). (b) Artificial detuning encoder ($f$): error in MHz for detunings in the 4--6~MHz range used to mimic experimental offset conditions. (c) Inhomogeneous dephasing-time encoder ($T_2^\ast$): error $\hat{T}_2^\ast - T_2^\ast$ (in $\mu$s) across $T_2^\ast\in[0.5,10]~\mu$s. }

\end{figure*}
\begin{table*}[t]
\caption{\label{tab:aux_sim_rmse_row}
\textbf{Simulation RMSE of auxiliary encoders}
RMSE is reported versus the discrete ground-truth label values for each auxiliary regressor; sample counts are omitted.}
\centering
\setlength{\tabcolsep}{4pt}
\renewcommand{\arraystretch}{1.15}
\scriptsize
\begin{ruledtabular}
\begin{tabular}{lcccccccccc}
\textbf{PL (\%)} &
75 & 80 & 85 \\
\textbf{RMSE (PL)} &
0.02964 & 0.10810 & 0.03391 \\
\hline
\textbf{$f$ (MHz)} &
4.0 & 4.5 & 5.0 & 5.5 & 6.0 \\
\textbf{RMSE (MHz)} &
0.08138 & 0.04827 & 0.03648 & 0.05016 & 0.08127 \\
\hline
\textbf{$T_2^\ast$ ($\mu$s)} &
0.5 & 1.0 & 1.5 & 2.0 & 2.5 \\
\textbf{RMSE ($\mu$s)} &
0.0311 & 0.0857 & 0.203 & 0.274 & 0.282 \\
\textbf{$T_2^\ast$ ($\mu$s)} &
3.0 & 3.5 & 4.0 & 4.5 & 5.0 \\
\textbf{RMSE ($\mu$s)} &
0.300 & 0.310 & 0.380 & 0.502 & 0.678 \\
\textbf{$T_2^\ast$ ($\mu$s)} &
5.5 & 6.0 & 6.5 & 7.0 & 7.5 \\
\textbf{RMSE ($\mu$s)} &
0.908 & 1.15 & 1.35 & 1.48 & 1.54 \\
\textbf{$T_2^\ast$ ($\mu$s)} &
8.0 & 8.5 & 9.0 & 9.5 & 10.0 \\
\textbf{RMSE ($\mu$s)} &
1.57 & 1.62 & 1.68 & 1.78 & 1.93 \\
\end{tabular}
\end{ruledtabular}
\end{table*}
\textbf{Simulation Results}
Figure~\ref{fig:auxiressim} and Table~\ref{tab:aux_sim_rmse_row} summarize simulation-only generalization of the three auxiliary ``oscillation-pattern'' encoders on the simulation test set. Across all three tasks, the prediction-error distributions are tightly centered near zero at each discrete label value, indicating that the encoders learn largely unbiased mappings under the same physical simulator and noise statistics used to generate the training corpus.

For the PL spin contrast encoder [Fig.~\ref{fig:auxiressim}(a)], the errors remain narrowly concentrated at all representative operating points, with RMSEs at the few-$10^{-2}$--$10^{-1}$ PL(\%) level (Table~\ref{tab:aux_sim_rmse_row}). This is consistent with contrast acting as a dominant, low-dimensional degree of freedom in the simulated traces: it primarily rescales the overall Ramsey response and is therefore strongly constrained by coarse waveform statistics, even when fast fluctuations are present. In the context of simulation-based hyperfine inference (Sec.~\ref{sec:hf_estimator}), providing an accurate contrast summary helps separate global amplitude variation from the weaker, phase-coherent beating signatures that carry hyperfine information.

The artificial detuning encoder [Fig.~\ref{fig:auxiressim}(b)] achieves sub-$0.1$~MHz RMSE throughout the 4--6~MHz range used in the simulator, with distributions concentrated near zero for each label. Within the synthetic pipeline, this accuracy directly supports the behavior seen in Fig.~\ref{sim-C13PRE}: because the hyperfine predictor must distinguish beating patterns rather than absorb global frequency shifts, an explicit detuning estimate reduces the effective nuisance variability in the time--frequency representation and makes the remaining classification ambiguity more clearly attributable to intrinsic spectral congestion at larger $n$.

The $T_2^\ast$ encoder [Fig.~\ref{fig:auxiressim}(c)] tracks inhomogeneous dephasing with small errors at short-to-intermediate $T_2^\ast$, while the spread grows toward longer $T_2^\ast$ values (Table~\ref{tab:aux_sim_rmse_row}). This trend is expected within the fixed simulation window: as $T_2^\ast$ increases, the envelope becomes progressively flatter over the sampled evolution times and is therefore less identifiable from the trace alone. Importantly, the error remains centered near zero across the full range, indicating that the encoder captures the correct monotonic relationship between envelope curvature and dephasing time. This behavior aligns with the hyperfine results in Fig.~\ref{sim-C13PRE}(b), where regression errors broaden primarily in regimes that are intrinsically less identifiable (large $n$ and congested spectra), rather than from systematic bias in these auxiliary summaries.

\textbf{Experimental Results}
\begin{figure}[t]
    \centering
    \includegraphics[width=1\linewidth]{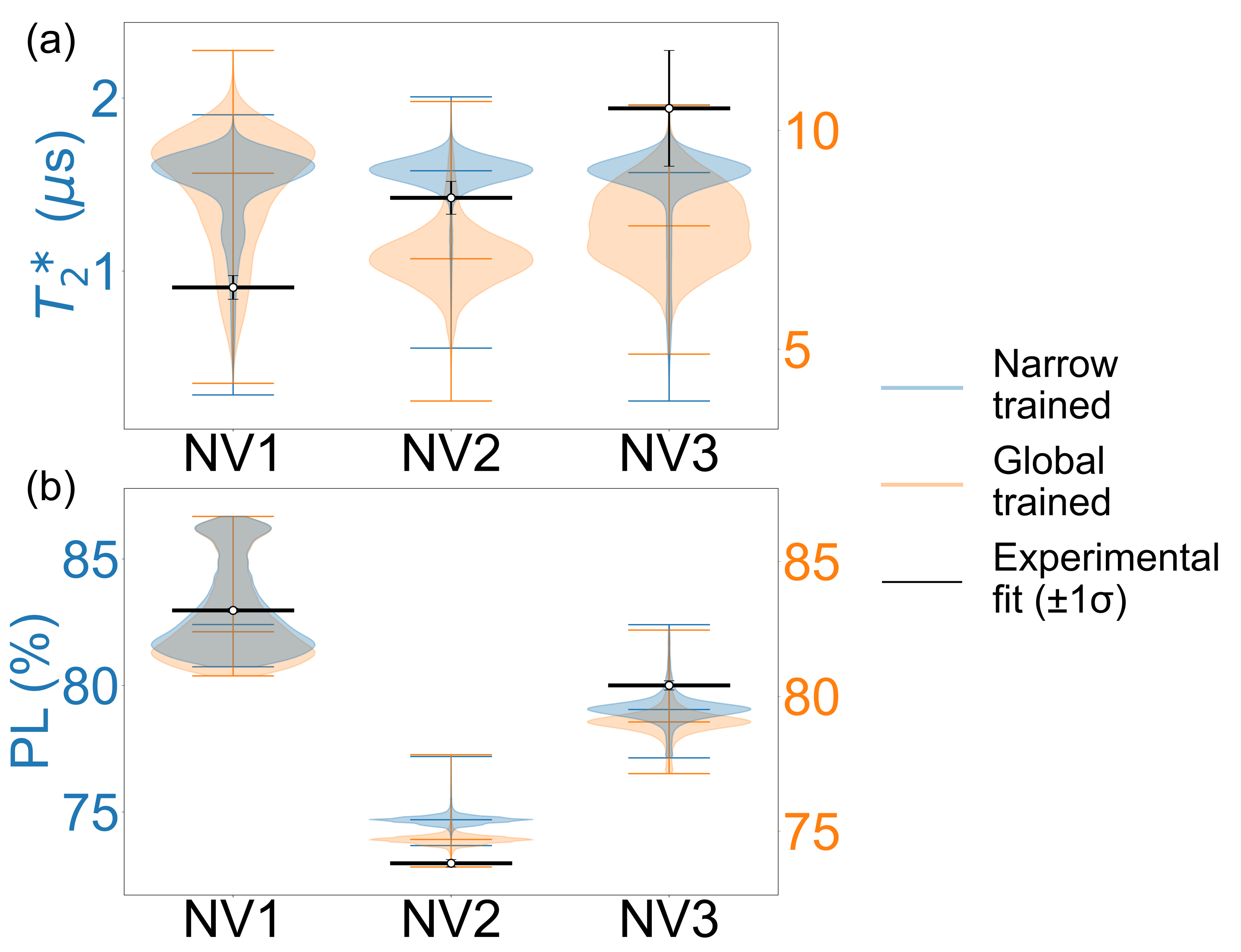}
    \caption{\label{fig:auxiresexp}
    \textbf{Experimental performance of the auxiliary $T_2^\ast$ and PL-contrast encoders.}
    Predictions are obtained from \emph{individual} few-shot experimental Ramsey traces (after denoising), while markers indicate independent high-SNR estimates from a dedicated experimental fit (mean $\pm 1\sigma$).
    (a) Predicted inhomogeneous dephasing time $T_2^\ast$ for NV1--NV3 using the \emph{same} predictor architecture trained under two dataset ranges:
    a narrow-range training set ($T_2^\ast \in [0.5,2]~\mu\mathrm{s}$; blue violins, left axis) and a global-range training set ($T_2^\ast \in [4,12]~\mu\mathrm{s}$; orange violins, right axis).
    Black markers show the experimental fit $T_2^\ast$($\pm 1\sigma$) value on left axis scale.
   (b) Predicted PL(\%) contrast for NV1--NV3 using the same two training ranges: a narrow-range trained encoder (blue violins, left axis) and a global-range trained encoder (orange violins, right axis). PL is reported as $100-\mathrm{Contrast}$ in percent units. Black markers denote the experimentally fitted PL($\pm 1\sigma$) values on left axis scale.
}
\end{figure}
Figure~\ref{fig:auxiresexp} evaluates the auxiliary encoders on held-out \emph{experimental} Ramsey traces and exposes a clear limitation of the $T_2^\ast$ predictor in its current sim-only form. In Fig.~\ref{fig:auxiresexp}(a), the globally trained model (synthetic $T_2^\ast=500$--$10000~\mathrm{ns}$; orange) performs poorly on these measurements, frequently producing multi-microsecond $T_2^\ast$ values that disagree with the independent multi-sweep fits, whereas training the \emph{same} architecture on a narrow-range dataset ($T_2^\ast=500$--$2000~\mathrm{ns}$; blue) substantially improves agreement and keeps predictions within the experimentally relevant regime (with residual bias remaining, most clearly for NV1 and NV3).We attribute this primarily to \emph{data-distribution mismatch}: under the present laboratory conditions the fitted $T_2^\ast$ values are concentrated in $\sim0.5$--$2~\mu\mathrm{s}$, so the experimental test set effectively constitutes a narrow-regime operating point that is marginal relative to the global-range training objective, leading to poor calibration in the region of interest.

Secondarily, the remaining discrepancies are consistent with a \emph{feature-extraction} mismatch between simulation and experiment.
Because the predictors operate on denoiser outputs rather than raw sweeps, small systematic distortions of the recovered envelope (amplitude, baseline, or normalization) can bias the scalar summaries extracted from the trace.
This mechanism naturally affects not only $T_2^\ast$ but also the PL-contrast encoder, and is consistent with the slight but systematic offsets in Fig.~\ref{fig:auxiresexp}(b) between the predicted $100-\mathrm{Contrast}$ distributions and the experimental fit markers. Notably, the contrast predictions exhibit substantially lower variance than $T_2^\ast$ across repeated few-shot traces, suggesting that contrast is an easier and more stable target for the encoder, but also making small systematic calibration errors easier to detect. In this sense, Fig.~\ref{fig:auxiresexp}(b) serves as a complementary diagnostic: it indicates that the learned features capture the correct relative contrast ordering across different NV centers, while still requiring modest experimental calibration for absolute accuracy.

In practice, these observations motivate a \emph{range-conditioned deployment} strategy: maintain a small library of checkpoints trained on different $T_2^\ast$ ranges and select (or ensemble) the checkpoint matched to the expected experimental regime, thereby preserving global coverage while improving local accuracy. To further reduce feature-level bias, one can introduce lightweight experimental adaptation (e.g., a small adapter layer or calibration head) trained on high-SNR experimental fits; however, obtaining sufficiently large, high-quality labeled experimental datasets is costly, motivating data-efficient adaptation schemes and careful prioritization of calibration measurements.

\section{Conclusion and Outlook}

In this work, we introduced \textsc{NVRNet}, a physics-informed deep-learning pipeline for rapid NV characterization in the few-sweep Ramsey regime. The method couples (i) a dual-path time--frequency denoising network pretrained on large-scale Hamiltonian-based simulations with experimentally calibrated noise statistics and (ii) lightweight, uncertainty-aware adapter modules fine-tuned on experimental traces to close the simulation-to-reality gap without retraining the full core. The denoised output is then fed to a hyperfine-parameter estimator trained on simulation labels, enabling direct prediction of the ${}^{13}\mathrm{C}$ count (up to $n_{\max}=9$) and parallel hyperfine couplings. Across held-out experimental data restricted to low averaging ($K\le 10$), the core+adapter model consistently improves over the raw baseline for three distinct NV centers, with substantial median RMSE reductions for NV~1 and NV~2 and a positive, albeit smaller, improvement for NV~3. Importantly, the simulation-only core model is not a reliable deployment strategy in this regime: it degrades severely for certain NVs and exhibits weak sensitivity to changes in experimental noise level, highlighting the practical necessity of targeted experimental adaptation. For hyperfine inference, forward reconstruction from predicted parameters reproduces the dominant time- and frequency-domain features of experimental traces, indicating that the estimator recovers physically meaningful spectral content that can serve as a principled initialization for conventional fitting workflows.

These results establish two practical conclusions. First, simulation pretraining alone is insufficient for robust few sweep denoising on experimental NV data even when synthetic noise is carefully engineered; lightweight adapter fine-tuning provides a high-leverage and parameter-efficient correction that materially improves the results. Second, physics-trained parameter prediction can reduce the human overhead of hyperfine model initialization by producing consistent, interpretable candidate parameters whose forward reconstructions agree with experimental FFT features, thereby accelerating the path from a short Ramsey acquisition to actionable hyperfine information.

Several extensions could further strengthen \textsc{NVRNet}. Simulation-to-reality robustness may be improved by incorporating additional experimental nonidealities—such as time-base jitter, pulse-area errors, slow frequency drift, and correlated readout fluctuations—either directly in the simulator or through learned noise models, and by making the denoiser more calibration-aware via uncertainty-aware training or explicit uncertainty prediction. For hyperfine estimation, spectral congestion at high ${}^{13}\mathrm{C}$ density motivates permutation-invariant or probabilistic outputs that represent ambiguity rather than single point estimates; performance may also improve as larger, well-curated experimental databases become available for training and benchmarking. More broadly, integrating the pipeline into closed-loop experiments could enable adaptive selection of Ramsey sampling grids and averaging strategies, while the same physics-pretraining–plus–adapter framework should transfer naturally to richer NV characterization tasks, including dynamical-decoupling fingerprints, multi-axis sensing, and register calibration.

\section{Acknowledgments}
This work was supported by Q-NEXT through the U.S. Department of Energy, Office of Science, National Quantum Information Science Research Centers. This work was performed in part at the Cornell NanoScale Facility, an NNCI member supported by NSF Grant NNCI-2025233.

\appendix
\section{Models Training Details}
We optimize all networks with AdamW~\cite{loshchilov2019decoupledweightdecayregularization}. For the UNet denoiser backbone, we use an initial learning rate of $5\times 10^{-4}$ on the full simulated dataset using a $0.7/0.3$ (train/validation) split. We employ a reduce-on-plateau learning-rate schedule: if the validation loss does not improve for five consecutive validation evaluations, the learning rate is decreased. We also use early stopping under the same patience criterion and retain the checkpoint with the lowest validation loss for all downstream experiments. Unless otherwise noted, core pretraining is run for 215 epochs (wall clock time $\sim 10$~h). For UNet Adapter fine-tuning, we minimizes the same composite objective used for core pretraining, evaluated on experimentally augmented batches (Appendix~\ref{DatAug}). Training is performed for up to 50 epochs (wall clock time$\sim 20$~min). All auxiliary encoders ($T_2^\ast$, overall contrast/PL, and detuning offset) and the hyperfine predictor are trained on the same fully simulated dataset used for denoiser backbone pretraining, for which the simulator can provides deterministic ground-truth labels ($T_2^\ast$, PL, detuning $f$, $^{13}$C spin count, and hyperfine parameters).  We report the checkpoint with the lowest validation loss. In our runs, the $T_2^\ast$ predictor converged after 35 epochs, the PL predictor after 22 epochs, the detuning-offset predictor after 56 epochs, and the hyperfine predictor after 50 epochs. All training is performed on a single NVIDIA RTX\,5090 GPU. 

\section{Details on lattice simulation and physical modeling}

\subsection{Lattice construction, cutoff, and ${}^{13}\mathrm{C}$ statistics}
\label{LatticeAppendix}

\paragraph*{Diamond supercell enumeration.}
We generate a finite diamond simulation lattice by explicitly enumerating a conventional-cell basis over an $n\times n\times n$ supercell. Using the lattice constant $a=3.57~\mathrm{\AA}$, atomic coordinates are constructed from integer ``mod-4'' basis points $\mathbf{p}_\mu$ (eight sites per conventional cell) and an FCC supercell index $\boldsymbol{\ell}=(i,j,k)\in\{0,\dots,n-1\}^3$ as
\begin{equation}
\mathbf{r}_{\ell,\mu}=\frac{a}{4}\left(\mathbf{p}_\mu + 4\,\boldsymbol{\ell}\right),
\qquad \mu=1,\dots,8.
\label{eq:lattice_enum}
\end{equation}
After enumeration, we recenter the coordinates by subtracting the geometric center of the bounding box, such that the simulation volume is symmetric about the origin. For the parameters used here ($n=5$), the integer coordinates span $0\le p_\mu + 4\ell \le 4n-1$ along each axis, so the resulting cubic box has side length
\begin{equation}
L=\frac{a}{4}(4n-1)\simeq 16.96~\mathrm{\AA},
\label{eq:lattice_L}
\end{equation}
which we refer to as a $\sim 15$--$17~\mathrm{\AA}$-scale ``large box'' used to seed dopant generation.

\paragraph*{NV placement and exclusion of first-shell carbons.}
An NV center is instantiated by selecting the lattice site closest to the origin as the vacancy and removing it from the coordinate list; the nearest neighbor to the vacancy is labeled as substitutional nitrogen. We then shift all coordinates so that the vacancy is at $\mathbf{r}=\mathbf{0}$. To avoid unphysical perturbations of the defect core, carbon sites in the first coordination shell of the vacancy are excluded from ${}^{13}\mathrm{C}$ substitution. In practice we identify this shell via a nearest-neighbor distance threshold
\begin{equation}
r_{\mathrm{nn}}=\frac{\sqrt{3}}{4}a,
\label{eq:nn_dist}
\end{equation}
and disallow substitution at sites with $r<r_{\mathrm{nn}}$ (up to a small numerical tolerance).

\paragraph*{Stochastic ${}^{13}\mathrm{C}$ substitution and dopant cap.}
To model the nuclear bath, each eligible carbon site $i$ is independently converted to ${}^{13}\mathrm{C}$ with probability $p_{13}=0.011$ (natural abundance), i.e.,
\begin{equation}
X_i\sim \mathrm{Bernoulli}(p_{13}),\qquad X_i=1 \Rightarrow i\in {}^{13}\mathrm{C},
\label{eq:bernoulli_doping}
\end{equation}
with independent draws across sites. To prevent rare high-density realizations from dominating memory and runtime, we also enforce a hard cap on the \emph{total} number of doped ${}^{13}\mathrm{C}$ sites generated in the large box (here $N_{{}^{13}\mathrm{C}}^{\max}=200$ in the generator).

\begin{figure}[tbp]
\centering
\includegraphics[width=0.7\linewidth]{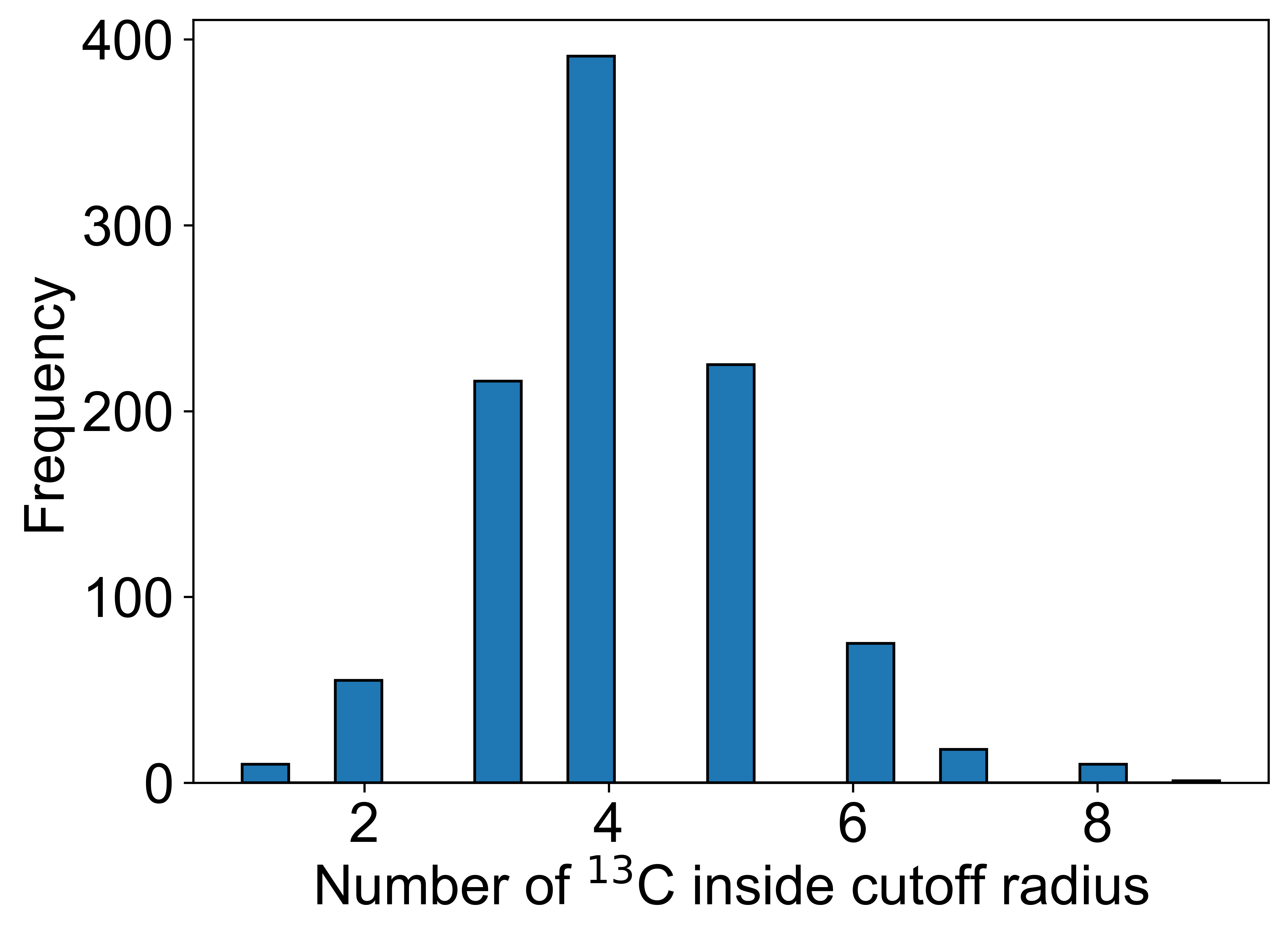}
\caption{\label{fig:configdis}
Distribution of retained ${}^{13}\mathrm{C}$ spins in the lattice-generated training set.}
\end{figure}

\paragraph*{Spherical cutoff and coupling-based justification.}
To bound the effective bath size while retaining only couplings that measurably modulate the Ramsey free-induction signal, we keep only ${}^{13}\mathrm{C}$ spins within a spherical cutoff radius $R_c$ from the NV center (here $R_c=6~\mathrm{\AA}$) and discard more distant dopants. For a nucleus at position $\mathbf{r}=(x,y,z)$ (NV axis $\parallel\hat{z}$), the lattice generator uses the secular dipolar scaling
\begin{equation}
C(\mathbf{r})=\frac{\alpha}{r^{3}}\left(3\frac{z^{2}}{r^{2}}-1\right)
=\frac{\alpha}{r^{3}}\left(3\cos^{2}\theta-1\right),
\label{eq:dipolar_cutoff}
\end{equation}
where $r=\|\mathbf{r}\|$, $\theta$ is the polar angle relative to the NV axis, and $\alpha$ collects constants and gyromagnetic ratios~\cite{abobeih2019atomic}. Using $\gamma_e\simeq 2\pi\times 2.8\times 10^{10}~\mathrm{rad}\,\mathrm{s}^{-1}\mathrm{T}^{-1}$ and $\gamma_{^{13}\mathrm{C}}\simeq 2\pi\times 1.07\times 10^{7}~\mathrm{rad}\,\mathrm{s}^{-1}\mathrm{T}^{-1}$~\cite{abobeih2019atomic}, the code conversion yields
\begin{equation}
\alpha \simeq 1.99\times 10^{4}~\mathrm{kHz}\cdot \mathrm{\AA}^{3}.
\label{eq:alpha_value}
\end{equation}
At the cutoff boundary $r=R_c=6~\mathrm{\AA}$, the characteristic scale is
\begin{equation}
\frac{\alpha}{R_c^{3}}\approx \frac{1.99\times 10^{4}}{6^{3}}~\mathrm{kHz}\approx 9.2\times 10^{1}~\mathrm{kHz},
\label{eq:coupling_scale_rc}
\end{equation}
and since $(3\cos^{2}\theta-1)\in[-1,2]$, the boundary couplings satisfy
\begin{equation}
|C(R_c,\theta)|\lesssim 2\,\frac{\alpha}{R_c^3}\approx 1.8\times 10^{2}~\mathrm{kHz}\ (\sim 0.18~\mathrm{MHz}).
\label{eq:coupling_bounds_rc}
\end{equation}
In our Ramsey acquisitions, spectral components far below this $\sim 0.1$--$0.2~\mathrm{MHz}$ scale produce oscillation periods comparable to (or longer than) the accessible evolution window and are further attenuated by the $T_2^\ast$ envelope, making them difficult to distinguish from slow baseline drift in few sweep traces. Because $|C|\propto r^{-3}$, increasing the cutoff rapidly admits many additional spins whose couplings fall below the experimentally resolvable bandwidth (e.g., at $r=8~\mathrm{\AA}$ the same scale drops to $\alpha/r^{3}\approx 3.9\times 10^{1}~\mathrm{kHz}$), increasing computational cost without commensurate gain in identifiable beating structure. We therefore adopt $R_c=6~\mathrm{\AA}$ together with the dopant cap to control worst-case bath size. The resulting distribution of retained ${}^{13}\mathrm{C}$ counts in the training set is shown in Fig.~\ref{fig:configdis}.

\subsection{\label{app:rwa_nv}Rotating-wave approximation for the NV center: elimination of perpendicular hyperfine terms and rotating-frame reduction}
This appendix derives the effective two-level rotating-frame Hamiltonian used in the Ramsey simulator and justifies neglecting perpendicular (spin-flip) hyperfine terms under our experimental conditions. In the following derivation, we treat $\hbar=1$ for convinience. We begin from the ground-state NV Hamiltonian (NV axis $\parallel \hat{z}$) including nearby nuclear spins,
\begin{equation}
H_{\mathrm{lab}} = H_{\mathrm{NV}}+H_{\mathrm{nuc}}+H_{\mathrm{hf}}+H_{\mathrm{MW}}(t),
\label{eq:app_Hlab}
\end{equation}
with
\begin{align}
H_{\mathrm{NV}} &= D S_z^2 + \gamma_e B_0 S_z,\\
H_{\mathrm{nuc}} &= \sum_{j}\gamma_{n,j} B_0 I_{z,j},\\
H_{\mathrm{hf}} &= \sum_{j}\mathbf{S}\cdot \mathbf{A}_j \cdot \mathbf{I}_j,
\end{align}
where $\frac{D}{2\pi} \simeq 2.87~\mathrm{GHz}$, $B_0$ is the static bias field aligned to $\hat{z}$, and $\mathbf{A}_j$ is the hyperfine tensor for nucleus $j$ (the intrinsic ${}^{14}$N and retained ${}^{13}$C spins). The microwave drive is modeled as a near-resonant transverse field,
\begin{equation}
H_{\mathrm{MW}}(t)=\Omega \cos(\omega_{\mathrm{MW}} t+\phi)\,S_x,
\label{eq:app_drive}
\end{equation}
where $\Omega$ is the Rabi frequency set by the applied microwave amplitude.

\paragraph*{Projection to the $\{|m_s=0\rangle,|m_s=-1\rangle\}$ manifold.}
In the bias-field regime of this work, the $|m_s=+1\rangle$ level is far detuned from the addressed $|0\rangle\leftrightarrow|-1\rangle$ transition and is not appreciably populated. We therefore restrict the electronic Hilbert space to the two-dimensional subspace
$\mathcal{H}_e=\mathrm{span}\{|0\rangle,|-1\rangle\}$.
Within $\mathcal{H}_e$, one may define Pauli operators
$\sigma_z = |0\rangle\langle 0| - |-1\rangle\langle -1|$,
$\sigma_+ = |0\rangle\langle -1|$, and $\sigma_- = |-1\rangle\langle 0|$.
Up to an irrelevant energy shift, the bare electronic Hamiltonian in this subspace is
\begin{equation}
H_{e}^{(2)} \equiv P_{\mathcal{H}_e}(H_{\mathrm{NV}})P_{\mathcal{H}_e}
= \frac{\hbar\omega_0}{2}\,\sigma_z,
\qquad
\omega_0 \approx D-\gamma_e B_0,
\label{eq:app_omega0}
\end{equation}
where $P_{\mathcal{H}_e}$ denotes projection onto $\mathcal{H}_e$. (The sign convention for $\omega_0$ depends on the choice of $\sigma_z$; only frequency differences matter below.)

\paragraph*{Parallel/perpendicular hyperfine decomposition.}
In the NV principal-axis frame, the hyperfine interaction for each nucleus can be decomposed into longitudinal (secular) and transverse (nonsecular) parts,
\begin{equation}
\begin{aligned}
H_{\mathrm{hf}} &= H_{\parallel}+H_{\perp},
\qquad 
\\
H_{\parallel}&=\sum_j A_{\parallel,j}\,S_z I_{z,j},
\qquad \\
H_{\perp}&=\sum_j \frac{A_{\perp,j}}{2}\left(S_+ I_{-,j}+S_- I_{+,j}\right),
\label{eq:app_hf_decomp}
\end{aligned}
\end{equation}
with $S_{\pm}=S_x\pm iS_y$ and $I_{\pm,j}=I_{x,j}\pm i I_{y,j}$ \cite{Childress2006Science,Doherty2013PhysRep,Awschalom2018NatPhoton}. The term $H_{\parallel}$ preserves the electronic spin projection and produces a state-dependent energy shift, while $H_{\perp}$ induces electron--nuclear flip-flops.

\paragraph*{Interaction picture with respect to the fast electronic splitting.}
To make the time-scale separation explicit, we move to the interaction picture generated by the dominant electronic term $H_e^{(2)}$ of Eq.~\eqref{eq:app_omega0}. For any operator $O$, define
$O^{(I)}(t)=e^{+i H_e^{(2)} t/\hbar} O\,e^{-i H_e^{(2)} t/\hbar}$.
Because $[H_e^{(2)},S_z]=0$, the longitudinal term is stationary,
\begin{equation}
H_{\parallel}^{(I)}(t)=H_{\parallel}.
\end{equation}
By contrast, the transverse electronic ladder operators rotate at $\omega_0$,
\begin{equation}
S_+^{(I)}(t)=e^{+i\omega_0 t}S_+,\qquad
S_-^{(I)}(t)=e^{-i\omega_0 t}S_-,
\label{eq:app_Spm_rot}
\end{equation}
so the transverse hyperfine interaction becomes
\begin{equation}
H_{\perp}^{(I)}(t)=
\sum_j \frac{A_{\perp,j}}{2}
\left(
e^{+i\omega_0 t}\,S_+ I_{-,j}
+
e^{-i\omega_0 t}\,S_- I_{+,j}
\right),
\label{eq:app_Hperp_I}
\end{equation}
where we have neglected the much slower nuclear Larmor rotation (typically MHz) relative to the GHz electronic splitting, since it does not change the averaging argument below.

For Ramsey evolution times $\tau$ satisfying $\omega_0 \tau \gg 1$ and couplings $A_{\perp,j}\ll \omega_0$ (our regime), the oscillatory terms in Eq.~\eqref{eq:app_Hperp_I} average to (near) zero at leading order. Formally, the first Magnus term for $H_{\perp}^{(I)}(t)$ scales as
\begin{equation}
\left\|\frac{1}{\tau}\int_0^{\tau} H_{\perp}^{(I)}(t)\,dt\right\|
\sim \mathcal{O}\!\left(\frac{A_{\perp}}{\omega_0\tau}\right)\ \rightarrow\ 0,
\end{equation}
so that $H_{\perp}$ does not contribute to the effective Hamiltonian to leading order \cite{abragam1961principles}. The resulting secular hyperfine Hamiltonian is therefore
\begin{equation}
H_{\mathrm{hf}}^{(\mathrm{sec})}\simeq H_{\parallel}=\sum_j A_{\parallel,j}\,S_z I_{z,j}.
\label{eq:app_secular}
\end{equation}
Residual effects of $H_{\perp}$ enter at higher order as small energy shifts scaling as $\sim A_{\perp}^2/\omega_0$ and are neglected in this work.

\paragraph*{Rotating frame at the microwave frequency (RWA for the drive).}
During the Ramsey pulses we apply the standard rotating-wave approximation to the microwave drive of Eq.~\eqref{eq:app_drive} by transforming to a frame rotating at $\omega_{\mathrm{MW}}$ in the $\{|0\rangle,|-1\rangle\}$ manifold. After discarding counter-rotating terms at $\sim 2\omega_{\mathrm{MW}}$, the driven two-level Hamiltonian takes the usual form
\begin{equation}
H_{\mathrm{rot}}(t)
\simeq
\frac{\Delta}{2}\sigma_z
+
\frac{\Omega}{2}\left(\cos\phi\,\sigma_x+\sin\phi\,\sigma_y\right),
\label{eq:app_Hrot_drive}
\end{equation}
where $\Delta$ is the detuning between the applied microwave and the (hyperfine-shifted) $|0\rangle\leftrightarrow|-1\rangle$ transition. In the Ramsey \emph{free-precession} interval the drive is off ($\Omega=0$), so only the detuning term remains.

\paragraph*{Effective detuning from quasi-static nuclear projections.}
Under the secular approximation of Eq.~\eqref{eq:app_secular}, the nuclei contribute additively as a longitudinal frequency shift of the electronic transition. Treating the nuclear bath as quasi-static over a single Ramsey shot, we replace each nuclear operator $I_{z,j}$ by its eigenvalue $m_j$. In our notation,
$m_N\in\{-1,0,+1\}$ for ${}^{14}$N and
$m_{C,i}\in\{-\tfrac12,+\tfrac12\}$ for each retained ${}^{13}$C. The resulting detuning used in simulation is
\begin{equation}
\begin{aligned}
\Delta &= \delta+ A_{\parallel}^{(N)} m_N + \sum_{i=1}^{N_C} A_{\parallel,i}^{(C)} m_{C,i},
\qquad \\
\end{aligned}
\label{eq:app_detuning}
\end{equation}
where $\delta$ is the offset detuning used to match experimental operating conditions and facilitate FFT-based parameter extraction (see Appendix~\ref{DatAug}). Equation~\eqref{eq:app_detuning} is the effective Hamiltonian input to our Ramsey simulator: during free evolution, the state accumulates a relative phase $\Delta\tau$ in the $\{|0\rangle,|-1\rangle\}$ manifold, generating the multi-frequency beating patterns later exploited for hyperfine-parameter estimation.

\paragraph*{Ramsey signal, ensemble averaging, and $T_2^\ast$ envelope.}
For a fixed quasi-static detuning $\Delta$ and ideal $\pi/2$--$\tau$--$\pi/2$ sequence, the population in $|0\rangle$ after the second pulse takes the standard form
\begin{equation}
P_0(\tau \,|\, \Delta)=\frac{1}{2}\left[1+\cos(\Delta \tau)\right],
\label{eq:app_P0_basic}
\end{equation}
up to an overall phase convention set by pulse axes. We model inhomogeneous dephasing by multiplying the oscillatory term by an empirical Gaussian envelope, $\exp[-(\tau/T_2^\ast)^2]$, and then ensemble-average over the relevant nuclear-spin projection combinations (all $m_N$ and $\{m_{C,i}\}$) to obtain the simulated Ramsey fringe used for training and evaluation. For the global training dataset, $T_2^\ast$ is sampled from $500$ to $10000~\mathrm{ns}$ in increments of $100~\mathrm{ns}$, while for the narrowed dataset it is limited to $500$--$2000~\mathrm{ns}$. In addition, to improve robustness against modest electronic noise and slow experimental drift, we simulate artificial detunings $\delta$, where $\delta=2\pi f_{d}$, where $f_{d} \in \{4.0,\,4.5,\,5.0,\,5.5,\,6.0\}~\mathrm{MHz}$, even though the nominal experimental value is $5~\mathrm{MHz}$.

\section{Experimental Data Acquisition, Augmentation and Uncertainty Propagation Methods}
\label{DatAug} 
\paragraph*{Experimental Acquisition and normalized-PL construction.}
\begin{figure}[t]
    \centering
    \includegraphics[width=\linewidth]{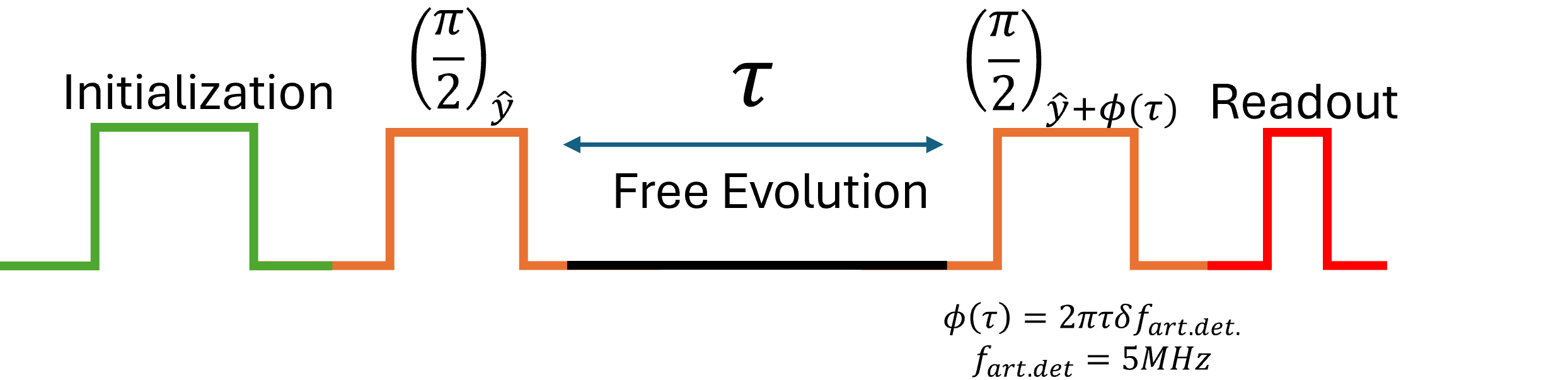}
    \caption{\label{fig:ramsey_sequence_appendix}
    \textbf{Ramsey pulse sequence with phase-ramped artificial detuning.}
    The spin is first optically initialized, followed by a $(\pi/2)_{\hat y}$ microwave pulse, a free-evolution interval $\tau$, and a second $\pi/2$ pulse before optical readout. In the Ramsey measurements reported here, the artificial detuning is by linearly advancing the phase of the second $\pi/2$ pulse as a function of $\tau$, such that $\phi(\tau)=2\pi \,f_{\mathrm{d}}\,\tau$ with $f_{\mathrm{d}}=5~\mathrm{MHz}$.}
\end{figure}
For each NV center, Ramsey measurements are acquired as repeated single-sweep experiments under nominally identical laboratory conditions. Each sweep uses the Ramsey microwave sequence shown in Fig.~\ref{fig:ramsey_sequence_appendix}, in which two $\pi/2$ pulses are separated by a variable free-precession interval and the artificial detuning ($f_{\mathrm{d}}=5~\mathrm{MHz}$) is introduced via a phase ramp on the second pulse. In a single sweep, the Ramsey sequence is evaluated on a discrete grid of 200 free-precession times $\{t_k\}_{k=1}^{200}$. At each time point $t_k$, the optical readout is integrated over two fixed temporal windows: a normalization window of width $1~\mu\mathrm{s}$ and a signal window of width $0.3~\mu\mathrm{s}$. Let $N(t_k)$ denote the photon counts collected in the normalization window and $S(t_k)$ the photon counts collected in the signal window. The normalized photoluminescence observable used throughout this work is then defined as
\begin{equation}
\mathrm{PL}(t_k)=100\times \frac{S(t_k)}{N(t_k)}.
\end{equation}
The factor of 100 expresses the readout on a percentage-scaled normalized-PL axis; it should therefore be interpreted as normalized PL readout rather than as a relative percent change.

To construct the high-SNR reference trace for a given NV, we acquire 200 such sweeps on the same time grid and under the same control settings. Each sweep produces a length-200 normalized PL trace. These traces are averaged pointwise across sweeps, and the resulting averaged curve is further fit by least squares to produce the reference trace $\mathrm{PL}_{\mathrm{ref}}(t)$. This fitted reference is used in the main text as the clean target for residual analysis and noise-model calibration.

The same acquisition convention is used for all experimental traces entering the dataset. In particular, the single-sweep traces are used to characterize residual noise statistics, while subsets of repeated sweeps are combined to emulate lower- or higher-averaging conditions for denoiser evaluation. Because both the signal and normalization channels are retained at the acquisition stage, the normalized PL observable and its associated uncertainty can be recomputed consistently for different sweep averages during downstream analysis.
\paragraph*{K-sweep averaging by random subset selection.}
Given the set of $N$ single-sweep traces, we form a synthetic $K$-sweep input by selecting, \emph{without replacement}, a subset of indices
\begin{equation}
\mathcal{I}_r^{(K)}=\{i_1,\dots,i_K\}\subset\{1,\dots,N\}, \qquad |\mathcal{I}_r^{(K)}|=K,
\end{equation}
and combining the corresponding sweeps into an averaged measurement. For each $K\in\{5,10,15,25,30,35,40,45,50\}$, we generate up to $N_{\mathrm{rep}}=10{,}000$ resampled realizations by drawing \emph{unique} index sets $\mathcal{I}_r^{(K)}$ using a fix ed random seed (for reproducibility). This procedure emulates the experimental practice of averaging multiple sweeps to improve signal-to-noise ratio (SNR), while producing a large, labeled test corpus for systematic evaluation.

\paragraph*{Photon-count aggregation and PL ratio.}
For each resampled realization $r$ and time sample $t_j$, we aggregate the photon counts from the normalization and signal windows,
\begin{equation}
N^{(r)}_j=\sum_{i\in\mathcal{I}_r^{(K)}} N_{i,j}, \qquad
S^{(r)}_j=\sum_{i\in\mathcal{I}_r^{(K)}} S_{i,j},
\end{equation}
where $N_{i,j}$ and $S_{i,j}$ denote the per-sweep normalization and signal counts (converted to raw counts from the stored kcounts units). The corresponding PL ratio trace is then computed from the sample means,
\begin{equation}
\begin{aligned}
\mathrm{PL}^{(r)}_j &= \frac{\mu_S(j)}{\mu_N(j)}, \qquad\\
\mu_S(j)&=\frac{1}{K}\sum_{i\in\mathcal{I}_r^{(K)}} S_{i,j}, \quad\\
\mu_N(j)&=\frac{1}{K}\sum_{i\in\mathcal{I}_r^{(K)}} N_{i,j}.
\end{aligned}
\end{equation}
In addition, we retain the simple summed PL channel
$\mathrm{PL}^{(r)}_{j,\mathrm{sum}}=\sum_{i\in\mathcal{I}_r^{(K)}} \mathrm{PL}_{i,j}$
for bookkeeping and sanity checks, although all quantitative evaluations in this work use the ratio-based PL representation.

\paragraph*{Uncertainty propagation.}
To propagate measurement uncertainty under finite $K$, we estimate the variance of the ratio-of-means using a first-order (delta-method) approximation that includes the empirical covariance between the signal and normalization channels across the selected sweeps. Let $\sigma_S^2(j)$ and $\sigma_N^2(j)$ be the sample variances of $\{S_{i,j}\}$ and $\{N_{i,j}\}$, and let $\mathrm{cov}_{SN}(j)$ be the sample covariance. The variance of $\mathrm{PL}^{(r)}_j=\mu_S(j)/\mu_N(j)$ is estimated as
\begin{equation}
\begin{aligned}
\mathrm{Var}\!\left[\mathrm{PL}^{(r)}_j\right]
&\approx \frac{1}{K}\Bigg[
\frac{\sigma_S^2(j)}{\mu_N^2( j)}
+ \frac{\mu_S^2(j)\,\sigma_N^2(j)}{\mu_N^4(j)} \\
&\qquad
- \frac{2\mu_S(j)\,\mathrm{cov}_{SN}(j)}{\mu_N^3(j)}
\Bigg]
\end{aligned}
\end{equation}
and we report the per-point standard uncertainty $u_{\mathrm{PL}}(j)=\sqrt{\max(\mathrm{Var},0)}$. This uncertainty is stored alongside each resampled trace and is used for downstream diagnostics and consistency checks.

\section{Interpretation of the Leading Principal Components}
\label{app:pca_interp}

\begin{figure}
\includegraphics[width=\linewidth,]{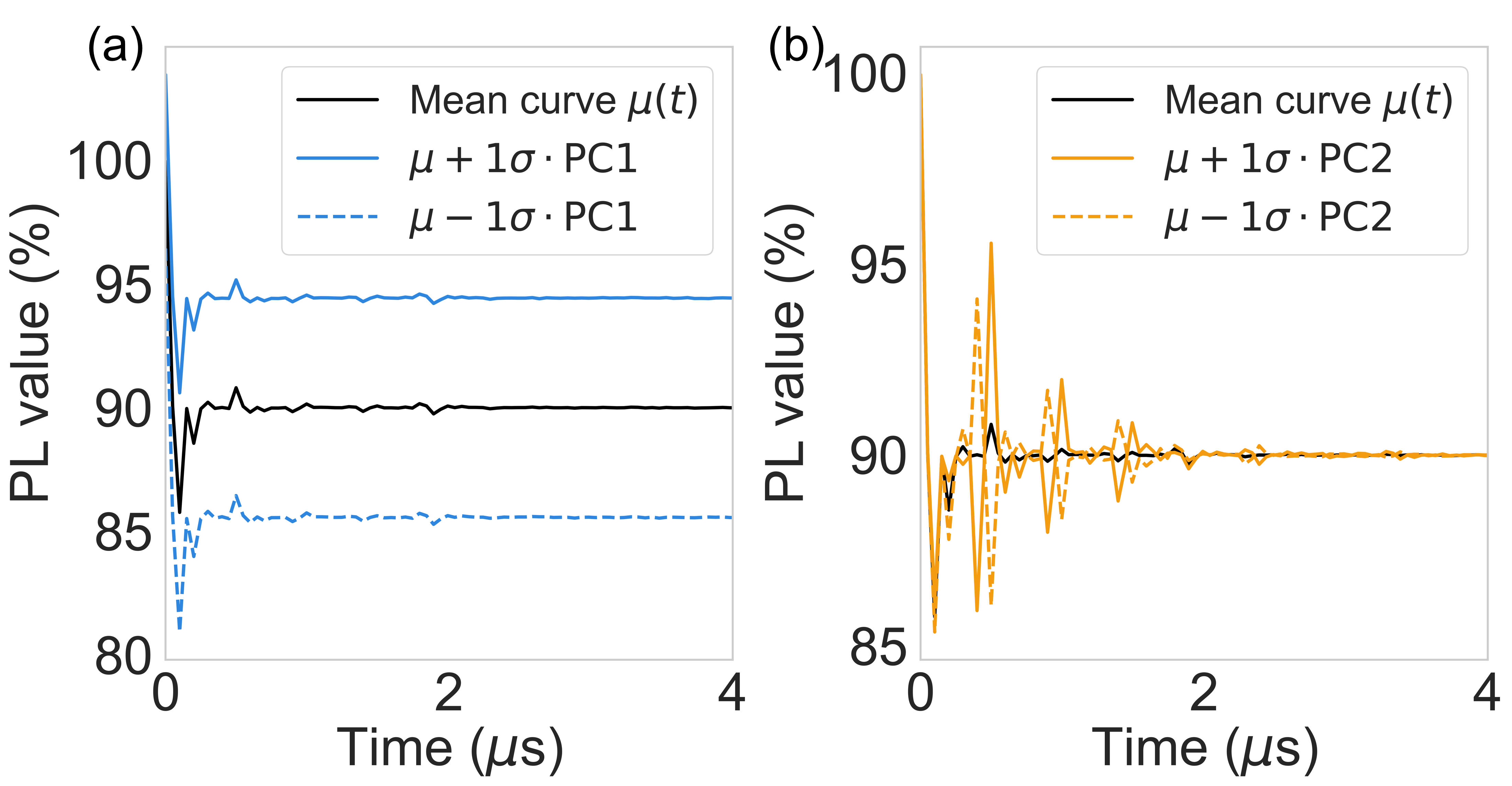}
\caption{\label{fig:PCAppendix}
\textbf{Projections of Leading Principal Components on Time Domain} Time-domain reconstructions obtained by projecting the mean trace onto $\pm$PC1 (a) and $\pm$PC2 (b), visualizing the primary modes of variation captured by the first two principal components.
}

\end{figure}
To interpret the leading principal components physically, we reconstruct time-domain modes around the empirical mean trace $\mu(t)$ using
\begin{equation}
x(t)\approx \mu(t)\pm \sigma_k u_k(t),
\end{equation}
where $u_k(t)$ is the $k$th principal component and $\sigma_k=\mathrm{std}(c_k)$ is the standard deviation of its score distribution. The corresponding reconstructions for the first two components are shown in Fig.~\ref{fig:PCAppendix}(a,b).

Figure~\ref{fig:PCAppendix}(a) indicates that $\mathrm{PC1}$ predominantly captures a global normalized-PL contrast or baseline variation: varying the score along this mode mainly shifts the overall level of the Ramsey trace while preserving its gross oscillatory structure. This is consistent with the experimentally observed spread in readout contrast, where the $m_s=0$ reference level lies near 100 in normalized-PL units and the $m_s=-1$ readout level typically falls in the $\sim 75$--$90$ range, corresponding to an overall contrast span of roughly 10--25 percentage points of normalized PL.

By contrast, Fig.~\ref{fig:PCAppendix}(b) shows that $\mathrm{PC2}$ primarily modulates the oscillatory structure of the trace rather than its global offset. In particular, motion along this component changes the relative beating pattern and local oscillation content, consistent with variations in hyperfine-induced frequency components arising from different nearby ${}^{13}\mathrm{C}$ configurations. In this sense, the first two principal components separate two experimentally relevant forms of variability: a dominant contrast/baseline degree of freedom (PC1) and a leading oscillatory-structure degree of freedom (PC2).
\section{Relation of RMSE to Residual-based Fitting Metrics}
\label{app:rmse_chi2}
The denoising metric used in the main text is the root-mean-square error (RMSE) between the denoised trace $\hat{y}(t)$ and the high-SNR reference trace $y(t)$. Writing the pointwise residuals as
\begin{equation}
r_i=\hat{y}(t_i)-y(t_i),
\end{equation}
one has
\begin{equation}
\mathrm{RMSE}^2=\frac{1}{T}\sum_{i=1}^{T} r_i^2,
\end{equation}
so the metric is simply the mean squared residual expressed in percentage points of normalized PL.

This quantity is the unweighted analogue of the residual measures that enter standard least-squares diagnostics. In a weighted analysis, one instead defines
\begin{equation}
\chi^2=\sum_{i=1}^{T}\frac{r_i^2}{\sigma_i^2},
\qquad
\chi^2_\nu=\frac{\chi^2}{\nu},
\end{equation}
with $\sigma_i$ the experimental uncertainty at time point $t_i$ and $\nu$ the number of degrees of freedom. Our use of RMSE avoids imposing a specific uncertainty model during denoiser evaluation, while preserving a direct interpretation as the typical residual mismatch to the experimental reference trace. Lower RMSE therefore implies improved compatibility with downstream residual-based fitting of $T_2^\ast$, detuning, and hyperfine-structure parameters.
\section{Deep Learning Model Structure details}
\subsection{Detailed architecture of the denoising network}
\label{app:denoise_details}

\paragraph*{Input representation.}
Each Ramsey trace consists of 200 uniformly sampled time points and is treated as a single-channel one-dimensional signal. During training and inference, traces are processed in mini-batches. For a batch of input traces, the time-domain input to the denoiser is therefore represented as a tensor of shape $(\mathrm{batchsize},1,200)$.

\paragraph*{Stage 1: frequency-domain coarse denoising.}
In the first stage, each input trace is transformed to the spectral domain using a real fast Fourier transform (rFFT). For a length-200 real-valued signal, this produces 101 complex frequency bins. We represent this complex spectrum using two real-valued channels corresponding to the real and imaginary parts, yielding an input tensor of shape $(\mathrm{batchsize},2,101)$.

The spectral tensor is then passed through a compact 1D UNet that performs coarse denoising directly in frequency space. The network outputs a denoised spectral representation with the same shape as the input. An inverse real Fourier transform maps this output back to the time domain, producing a coarse waveform reconstruction of shape $(\mathrm{batchsize},1,200)$. This stage is designed primarily to suppress broadband noise while retaining the narrowband spectral content associated with Ramsey beating.

\paragraph*{Stage 2: time-domain refinement.}
The coarse time-domain reconstruction from Stage 1 is then used as the input to a second denoising network operating directly in the time domain. This stage performs waveform-to-waveform refinement, removing residual local artifacts that remain after the spectral pass and improving recovery of the oscillatory envelope. In particular, this refinement stage helps preserve weak phase-coherent modulations that may be partially distorted by frequency-domain denoising alone.

\subsection{\label{app:unet_attention}Multi-head self-attention used at the UNet bottlenecks}
At the bottleneck of each 1D UNet we apply standard multi-head self-attention (MHA) \cite{vaswani2017attention}. Let $\mathbf{H}\in\mathbb{R}^{\text{batchsize}\times L_b\times d}$ denote the bottleneck representation written as a length-$L_b$ sequence of $d$-dimensional tokens. MHA with $H$ heads computes queries, keys, and values by learned linear projections
\begin{equation}
\mathbf{Q}=\mathbf{H}\mathbf{W}_Q,\qquad
\mathbf{K}=\mathbf{H}\mathbf{W}_K,\qquad
\mathbf{V}=\mathbf{H}\mathbf{W}_V,
\end{equation}
and evaluates scaled dot-product attention per head,
\begin{equation}
\mathbf{A}^{(h)}=\mathrm{softmax}\!\left(\frac{\mathbf{Q}^{(h)}\big(\mathbf{K}^{(h)}\big)^{\!\top}}{\sqrt{d_h}}\right),
\qquad
\mathbf{O}^{(h)}=\mathbf{A}^{(h)}\mathbf{V}^{(h)},
\end{equation}
where $d_h=d/H$ and the softmax is applied row-wise \cite{vaswani2017attention}. Head outputs are concatenated and projected,
\begin{equation}
\mathrm{MHA}(\mathbf{H})=\mathrm{Concat}\!\left(\mathbf{O}^{(1)},\dots,\mathbf{O}^{(H)}\right)\mathbf{W}_O,
\end{equation}
followed by a residual connection and layer normalization \cite{He2016ResNet,ba2016layer},
\begin{equation}
\mathbf{H}'=\mathrm{LayerNorm}\!\big(\mathbf{H}+\mathrm{MHA}(\mathbf{H})\big).
\end{equation}
In our implementation we place this block only at the bottleneck (rather than at full resolution), which preserves the UNet inductive bias while enabling efficient non-local interactions; related design choices have been explored in attention-augmented UNet variants \cite{Chen2021TransUNet,Hatamizadeh2022UNETR}.

\subsection{\label{app:hf_tokens}Token construction, embeddings, and Transformer self-attention for the hyperfine predictor}
This appendix specifies the exact tokenization and Transformer operations used by the hyperfine frequency predictor (Transformer head) in Sec.~\ref{sec:hf_estimator}.

\paragraph*{Per-trace normalization.}
For a single trace $\mathbf{y}\in\mathbb{R}^{200}$ we compute
\begin{equation}
\begin{aligned}
\widetilde{\mathbf{y}}&=\frac{\mathbf{y}-\mu(\mathbf{y})}{\sigma(\mathbf{y})+\epsilon},
\qquad\\
\mu(\mathbf{y})&=\frac{1}{200}\sum_{\ell=1}^{200}y_\ell,\quad\\
\sigma^2(\mathbf{y})&=\frac{1}{200}\sum_{\ell=1}^{200}\big(y_\ell-\mu(\mathbf{y})\big)^2,
\end{aligned}
\end{equation}
with a small $\epsilon>0$ for numerical stability.

\paragraph*{Time and frequency tokens.}
The \emph{time tokens} are the $200$ scalar samples $\{\widetilde{y}_\ell\}_{\ell=1}^{200}$. The \emph{frequency tokens} are derived from the real FFT of the normalized trace. Let
\begin{equation}
\begin{aligned}
X_k &=\mathrm{rFFT}(\widetilde{\mathbf{y}})_k,\qquad\\
s_k &=\log\!\big(1+|X_k|\big),k=1,\dots,101,
\end{aligned}
\end{equation}
so that $\{s_k\}_{k=1}^{101}$ provides a log-magnitude summary of the 101 real-FFT bins for a length-200 sequence.

\paragraph*{Metadata tokens and embedding dimension.}
We prepend (i) a learnable classification token and (ii) scalar metadata tokens encoding global waveform factors (contrast/PL, $T_2^\ast$, and optionally the offset detuning $f$ if used as an input token). Scalars are rescaled to comparable numeric ranges, e.g.
\begin{equation}
\mathrm{PL}_{\mathrm{sc}}=\frac{\mathrm{PL}}{100},\qquad
(T_2^\ast)_{\mathrm{sc}}=\log\!\big(1+T_2^\ast\big),
\end{equation}
(and similarly for $f$ if included). Each scalar token is embedded into the common model dimension $D=256$ via a learned linear map. The full token sequence length is
\begin{equation}
T_{\mathrm{tot}} = 1 + N_{\mathrm{meta}} + 200 + 101,
\end{equation}
where $N_{\mathrm{meta}}$ is the number of scalar metadata tokens used (typically $N_{\mathrm{meta}}=2$ for PL and $T_2^\ast$, or $3$ if $f$ is also included). The embedded input is then
\begin{equation}
\mathbf{E}\in\mathbb{R}^{\text{batchsize}\times T_{\mathrm{tot}}\times 256}.
\end{equation}
We add learned token-type embeddings (to distinguish classification, metadata, time, and frequency tokens) and learned positional embeddings along the $T_{\mathrm{tot}}$ sequence.

\paragraph*{Transformer encoder and self-attention.}
The embedded sequence is processed by a Transformer encoder \cite{vaswani2017attention}. For a layer input $\mathbf{H}\in\mathbb{R}^{\text{batchsize}\times T_{\mathrm{tot}}\times 256}$, multi-head self-attention forms queries, keys, and values by learned projections
\begin{equation}
\mathbf{Q}=\mathbf{H}\mathbf{W}_Q,\qquad
\mathbf{K}=\mathbf{H}\mathbf{W}_K,\qquad
\mathbf{V}=\mathbf{H}\mathbf{W}_V,
\end{equation}
and computes (for each head) scaled dot-product attention followed by head concatenation and an output projection:
\begin{equation}
\mathrm{MHA}(\mathbf{H})=\mathrm{Concat}\!\left(\mathbf{O}^{(1)},\dots,\mathbf{O}^{(H)}\right)\mathbf{W}_O.
\end{equation}
We use the standard pre-normalization encoder form (LayerNorm before attention and before the feed-forward sublayer), which improves optimization stability for moderate depth. The sequence-level representation is taken as the \emph{final} classification embedding \cite{Devlin2019BERT,Dosovitskiy2021ViT,Touvron2021DeiT},
\begin{equation}
\mathbf{h}_{\mathrm{cls}}\in\mathbb{R}^{\text{batchsize}\times 256},
\end{equation}
and is fed to the count and hyperfine regression heads described in Sec.~\ref{sec:hf_estimator}.

\section{Supplementary Results}

\subsection{False-Positive Test on the denoised result}
\begin{figure*}[t!]
\centering
\includegraphics[width=7 in]{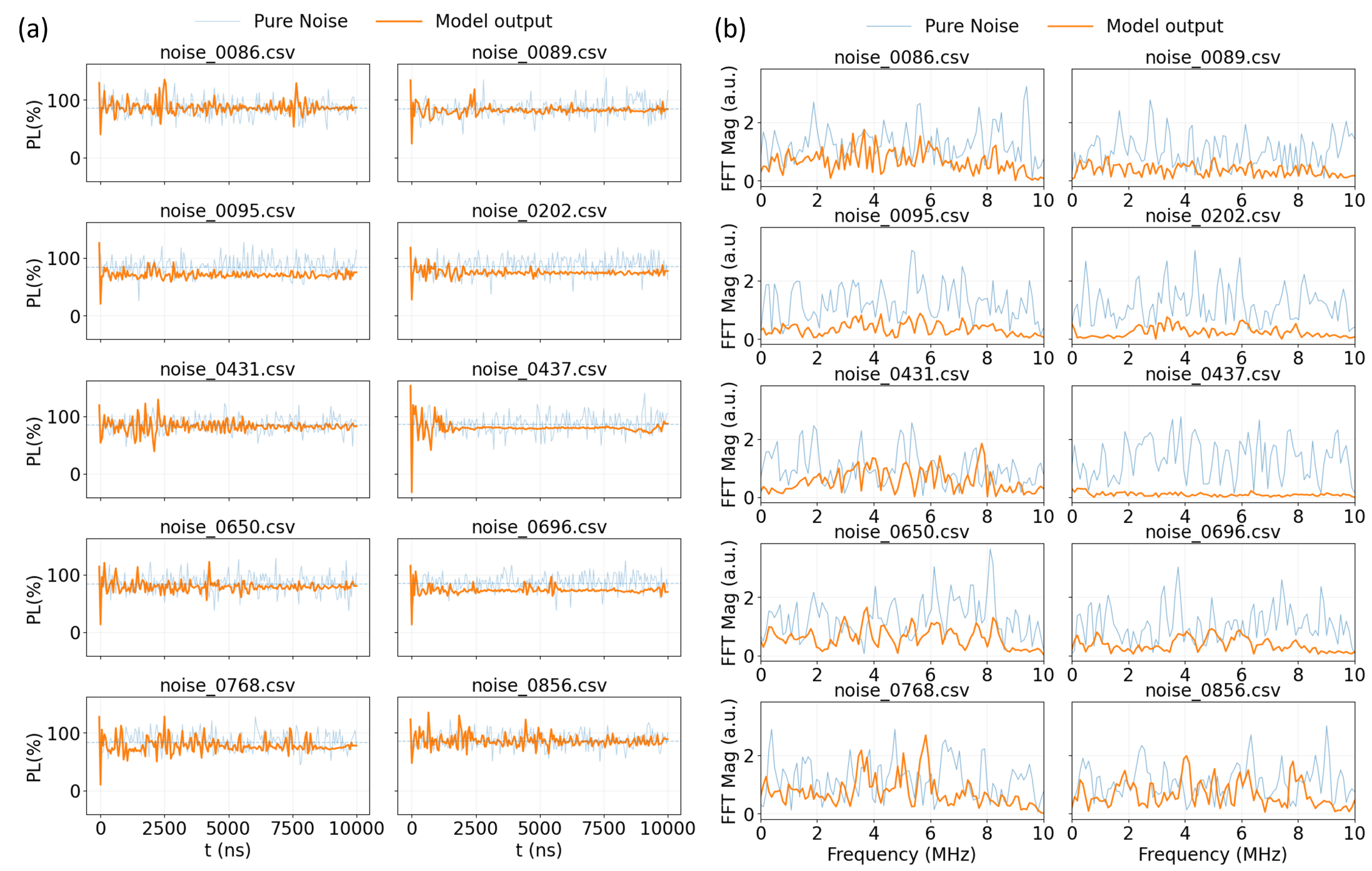}
\caption{\label{fig:false_positive}
\textbf{False-positive control using pure-noise inputs.}Ten randomly selected traces containing only noise are passed through the model. 
(a) Time-domain outputs: the network maps each noise trace (blue) to a smooth, near-constant prediction (orange) that closely follows the trivial baseline given by the trace mean (dashed), indicating no spurious Ramsey-like oscillations are generated. 
(b) Frequency-domain check: the corresponding rFFT magnitude spectra show that the model output suppresses broadband noise components and does not introduce structured peaks characteristic of hyperfine-induced beating. 
}

\end{figure*}

To verify that the denoiser does not hallucinate Ramsey-like structure when no physical signal is present, we perform a false-positive control using \emph{pure-noise} inputs. Specifically, we generate random traces with the same length and scale as the experimental PL readout and provide an uncertainty channel set pointwise as $u(t)=\sqrt{x(t)}$, mirroring the shot-noise scaling used elsewhere in our pipeline. These noise-only traces are then passed through the full denoising model (core+adapters). Figure~\ref{fig:false_positive} shows that most of the network outputs the noise signals, rather than producing oscillatory envelopes or beating patterns that could be misinterpreted as hyperfine signatures. The corresponding frequency-domain check further supports this conclusion: the rFFT magnitudes of the model outputs exhibit strong suppression of broadband components without introducing narrowband peaks. Together, these results indicate that the denoiser is not simply a generic oscillation generator; instead, it remains conservative on unstructured inputs and preferentially preserves oscillatory content only when it is supported by coherent structure in the input data.

\subsection{More result on Experimental Hyperfine Prediction and reconstruction}
\begin{figure*}[t!]
\centering
\includegraphics[width=7 in]{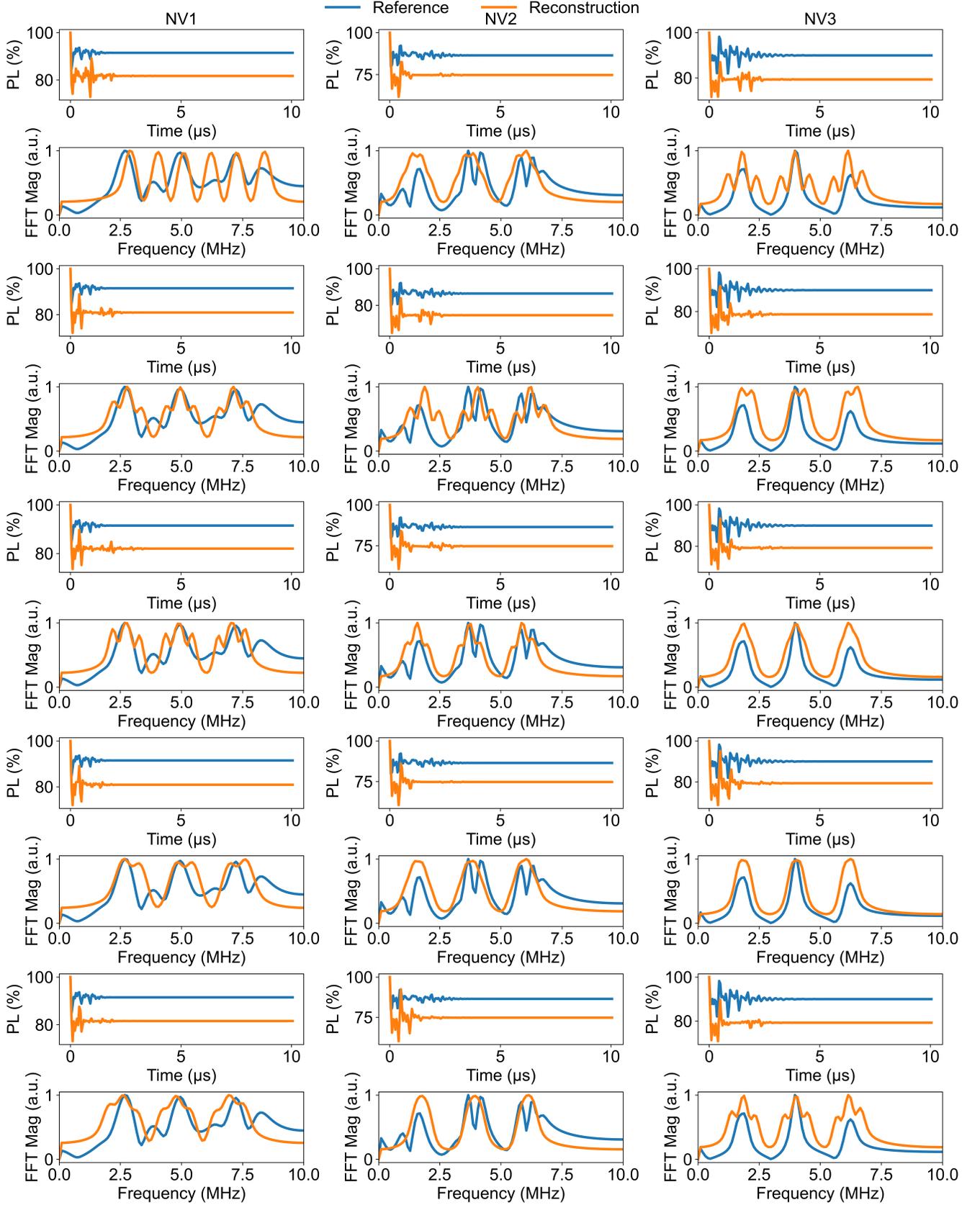}
\caption{\label{fig:morecomprehensHF}
\textbf{Additional examples of experimental hyperfine prediction via Ramsey-trace reconstruction.}
For each NV (NV~1--3), five reconstructions are randomly sampled from the held-out experimental test set. In each row group, the top panel shows the time-domain Ramsey PL(\%) trace (reference versus reconstruction), and the bottom panel shows the corresponding normalized FFT magnitude, demonstrating agreement of the dominant spectral bands across a range of experimental instances.}
\end{figure*}

To complement the representative examples shown in the main text (Fig.~\ref{fig:exp_hf_recon}), we provide additional qualitative results illustrating the consistency of the hyperfine-parameter estimator across the held-out experimental test set. Figure~\ref{fig:morecomprehensHF} shows multiple randomly selected reconstructions for each NV center, where predicted hyperfine parameters (obtained from the denoised traces) are used to forward reconstruct the Ramsey response. Across these examples, the reconstructions reproduce the dominant oscillatory content and the main frequency-band structure observed in the reference traces, supporting the conclusion that the estimator recovers physically meaningful hyperfine information.

\clearpage

\bibliography{apssamp}

\end{document}